\documentclass[times,onecolumn,preprint,authoryear]{elsarticle}
\usepackage{jasr}
\usepackage{framed,multirow}

\usepackage{amssymb}
\usepackage{latexsym}

\usepackage[switch]{lineno}

\usepackage{url}
\usepackage{xcolor}
\definecolor{newcolor}{rgb}{.8,.349,.1}

\usepackage[citebordercolor=white]{hyperref}

\journal{Advances in Space Research}

\begin{document}

\verso{Strauss \textit{etal}}

\begin{frontmatter}

\title{The updated SANAE neutron monitor}

%% or include affiliations in footnotes:
\author[NWU]{R.~D.~Strauss, dutoit.strauss@nwu.ac.za}

\author[NWU]{C.~van~der~Merwe, cobus.vandermerwe29@gmail.com}

\author[NWU]{C.~Diedericks, 27863840@nwu.ac.za}

\author[NWU]{H.~Kr\"uger, helenakruger6@gmail.com}

\author[NWU]{H.~G.~Kr\"uger, hendrik.kriiger@gmail.com}

\author[NWU]{K.~D.~Moloto, katlego.moloto@nwu.ac.za}

\author[sansa]{S.~Lotz, slotz@sansa.org.za}

\author[NWU]{G.~M.~Mosotho, mosothomoshegodfrey@gmail.com}

\address[NWU]{Center for Space Research, North-West University, Potchefstroom Campus, Private Bag X6001, 2520 Potchefstroom, South Africa}

\address[sansa]{South African National Space Agency, PO BOX 32, 7200 Hermanus, South Africa}

\begin{abstract}
Neutron monitors have been the premier ground-based instruments for monitoring the near-Earth cosmic ray flux for more than 70 years. It is essential to continue with such measurements in order to extend this unique long-term time series. Moreover, with the recent interest of the aviation industry to space weather effects, and especially the radiation risk posed by solar energetic particles and galactic cosmic rays, it is vital to extend the current neutron monitor network in order to provide near-real-time measurements to the space weather community. In this paper we discuss a new electronics system that was retrofitted to the SANAE neutron monitor in Antarctica. We present initial results from this system, featuring very high temporal resolution and discuss the techniques applied to the data analysis. Based on these successful upgrades, we are confident that this system can be used to rejuvenate the aligning neutron monitor network, and even possibly to revive some of the decommissioned instruments.
\end{abstract}

\begin{keyword}
cosmic rays \sep neutron monitor \sep space weather
\end{keyword}

\end{frontmatter}

%-------------------------------------------------------------------

\begin{figure}[!t]
\begin{center}
\noindent\includegraphics[width=0.55\textwidth]{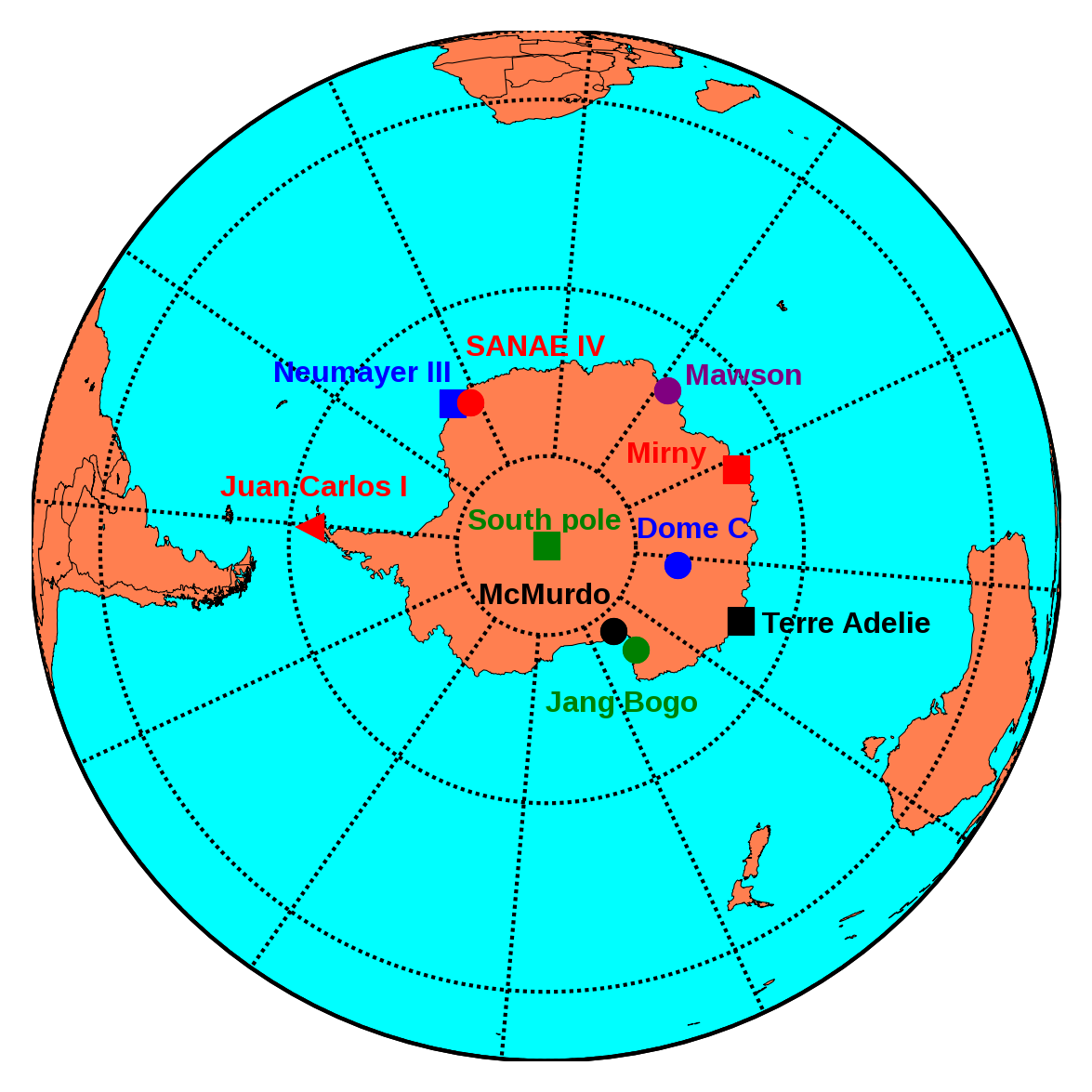}\\
\noindent\includegraphics[width=0.75\textwidth]{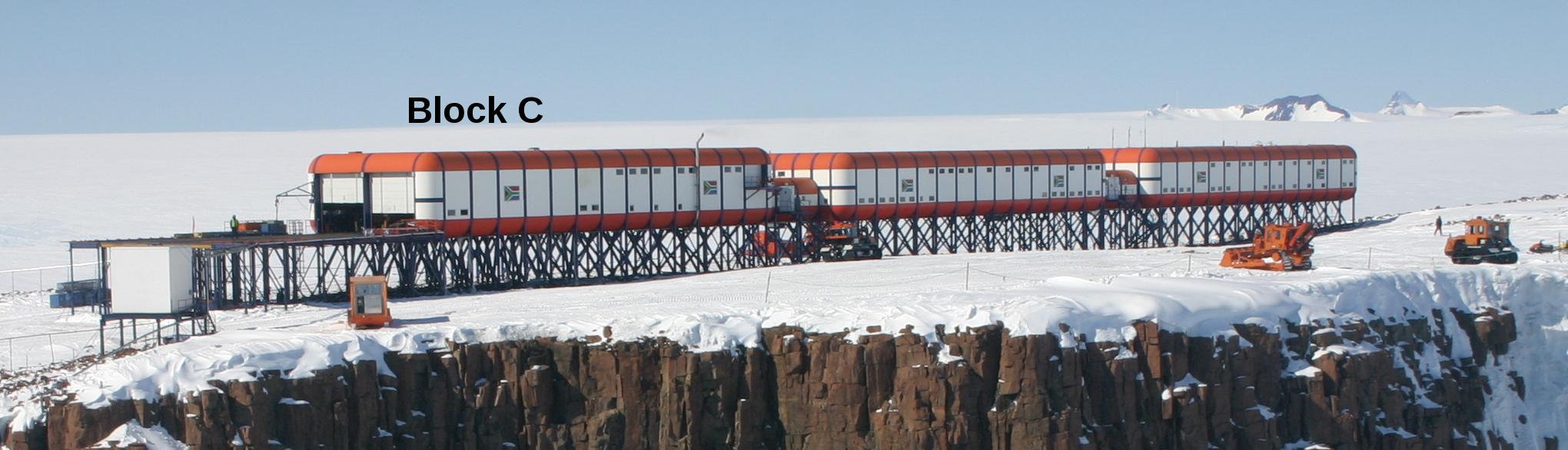}\\
\noindent\includegraphics[width=0.75\textwidth]{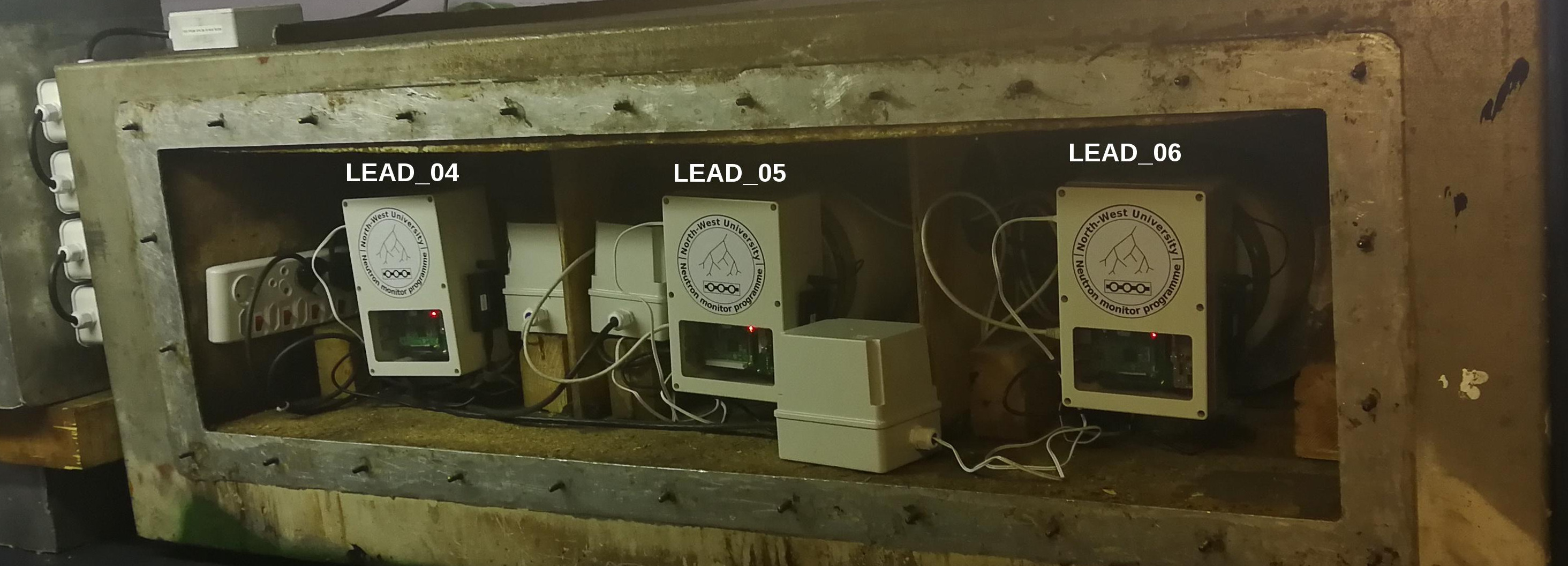}
\end{center}
\caption{The top panel shows the location of the SANAE IV base, along with a selection of other Antarctic bases operating NMs. The  SANAE NM, with its front cover removed, is shown in the bottom panel. The middle panel, showing the SANAE IV base, is a modified version of an image by Dr Ross Hofmeyr (CC BY-SA 3.0)}
\label{Fig:sanae_jpg_pic}
\end{figure}

\section{Introduction}

Cosmic rays, of galactic and solar origin, are the main source of harmful natural radiation at, and above, aviation altitudes. As such, their intensities must be monitored in near-real-time and the most cost effective instrument to do so, is the neutron monitor (NM). The near-real-time data from the South African NM network is available at \url{https://fskbhe1.puk.ac.za/neutronmonitor/}, and can be monitored live at \url{https://fskbhe1.puk.ac.za/spaceweather/}. Of course, NM observations are also used for theoretical studies, especially long-term cosmic ray studies, where the quality of the data, and the stability of the monitor, are  major concerns.\\

CRs can enter the Earth's atmosphere if they have sufficient energy to pass through the Earth's geomagnetic field. This minimum energy is given in terms of a cut-off rigidity (i.e. momentum per unit change), $P_c$, which depends strongly on geomagnetic latitude \citep[e.g.][]{dorman2004,Desorgher-etal-2009}. When CRs enter the atmosphere, they interact {with nuclei of atmospheric gases}, leading to showers of secondary particles. These secondary protons and neutrons are then detected on ground level by NMs.\\

These instruments, therefore, do not measure the cosmic radiation directly, but rather the flux of secondary and tertiary neutrons produced in the atmosphere, and in the detector itself. The count rate of a NM can thus be written, in the most basic formulation, as

\begin{equation}
    N(P_c,t) = \sum_i \int_{P_c}^{\infty} j_i (P,t) Y_i (P,t, \ldots) dP
\end{equation}

where $i$ represents the particle distribution under consideration (e.g. protons and helium particles), {$j_i (P,t)$ is the differential rigidity spectrum of cosmic rays of type $i$}, and $Y_i (P,t, \ldots)$ is the so-called yield function that represents the response of the instrument to the flux of primary cosmic rays. This parameter includes atmospheric and instrumental effects and therefore it depends on various parameters uniquely to each monitor \citep{clemdorman2000,rogelio2016,mishevetal2020}\\

The basic design of the NM has not changed much in 70 years \citep{simpson2000,Butikofer2018}. The main part of the instrument is the lead producer where secondary nucleons interact with lead nuclei, which in turn releases a number of {\it evaporated neutrons} \citep{bieberetal2004}. These suprathermal neutrons moves through a moderator (made from a proton-rich material; usually polyethylene in newer NMs, and paraffin wax in older types NMs), thermalizing their energies, and thus leading to a larger cross section to react with the BF$_3$ gas in the detector tube. There, the following neutron capture process takes place \citep{knoll12010}

\begin{eqnarray}
\label{Eq:reactions}
^{10}\mathrm{B} + \mathrm{n} & \rightarrow & \left\{ \begin{array}{ccc}
     ^{7}\mathrm{Li}\,\,\, + \alpha + 2.78 \, \mathrm{MeV} \\
     ^{7}\mathrm{Li}^{*} + \alpha + 2.30 \, \mathrm{MeV} \end{array} \right. 
\end{eqnarray}

where the $\alpha$ particle ionizes the gas as it is accelerated by an applied voltage, leading to a slight decrease in the applied voltage, and which is registered by appropriate electronics as a pulse. Note that about 6\% of the reactions leave $^7$Li$^*$ in an excited state. {$^3$He gas is also an efficient neutron capture gas and is used in some detectors.} Lastly, the NM is surrounded by a reflector (made from the same material as the moderator), which stops thermal neutrons from entering the monitor, while also trapping neutrons, produced in the lead producer, inside of the NM. Neutrons produced inside the producer from high energy protons tend to show a high level of multiplicity as a single cosmic ray proton {can produce multiple neutrons}. However, suprathermal neutrons, produced in the atmosphere, {and neutrons formed in the producer by low energy protons are} registered through the same neutron capture process, resulting in a low level of multiplicity. We will show how to differentiate between these two populations in a later section.\\

South Africa has had a permanent presence on the continent of Antarctica since 1959/1960 when the first overwintering team, South African National Antarctic Expedition (SANAE) 1, spent the winter in an abandoned Norwegian base. The first South African base, SANAE I (note the Roman numeral), was opened in 1962 during the SANAE 2 expedition. The SANAE I base was constructed on the Fimbulisen Ice Shelf in the Dronning Maud Land, located at 70$^{\circ}$18'S 2$^{\circ}$22'W. A 3NM64 (3 tubes of the 1964 super neutron monitor design) was installed at the SANAE I based by the SANAE 5 expedition in 1963/1964. The SANAE I base, however, suffered from snow accumulation, and the whole NM structure had to be raised above the new snow level each year. The same design was employed for subsequent bases, SANAE II (opened in 1971) and III (opened in 1979) at the same location, where after it was decided to design and build a more permanent structure. The SANAE IV base was opened in 1997 on the nunatak (a rocky outcrop) Vesleskarvet located at 72$^{\circ}$40'22"S 2$^{\circ}$50'26"W. During the move to the SANAE IV base, the NM was upgraded to a 6NM64.The top panel of Fig. \ref{Fig:sanae_jpg_pic} shows the location of SANAE IV, together with {a selection} {of} contemporary Antarctic research stations that, at some point, operated a NM. {This selection is by no means complete and does not included, for example, the new Syowa station \citep[][]{Katoetal2001}.} The middle panel of the figure shows the SANAE IV base, emphasizing ``Block C" which houses the NM, shown in the bottom panel with its front cover removed to show the newly installed electronics. The NM is located  on an elavated steel platform close to the building's roof. A historical account of the SANAE NM is given by \citet{moraaletal2011}.\\

In this paper we describe how the old NM tubes were retrofitted in February 2019 with a newly developed data acquisition system, which is a self-contained unit, featuring an integrated high-voltage (HV) supply, and a temporal resolution of 500ns. We describe the data reduction applied to the measurements, and how this new system can be used to track the integrity of the measurement through new proxies. We propose that the system described here can also be used to rejuvenate and/or restart older NM stations that were decommissioned due to either financial constraints or lack of replacement parts.

%--------------------------------------------------------------------
\section{New SANAE installation}

Although the SANAE IV NM installation was more environmentally stable than previous SANAE NM incarnations, there was still a major problem with noise in the high-voltage (HV) supply during windy conditions due to static build-up. Together with the ailing 50 year old electronics, is was decided to perform a complete update, beginning at the end of 2018. We decided to use the new electronic heads, developed for our mini-NM project, and described in detail by \citet{Straussetal2020}. \\

%--------------------------------------------------------------------
\subsection{Hardware changes}

The SANAE BP28 NM tubes are BP28F type \citep{fowler1963} used extensively in the International Quiet Sun Year (IQSY) NMs, also know as ``super-NMs", in the 1960's \citep{HattonCarmicheal1964}. All older electronics, including the attached pre-amplifiers, were completely replaced. The polarity of the tubes were also changed to a positive {central wire} and grounding the walls of the tubes to both the HV generator, but also the main ground pin of the base. 
The BP28's guard rings, which surrounds the {central wire} near the walls of the tube, were connected to the {central wire} through a 5$\Omega$ resistor \citep{Krugeretal2017}. The {central wire} and walls were connected to a normal HV connector, and the electronics head used in the mini-NMs were screwed onto each tube. The tubes are operated at a potential difference of 2850V \citep{fowler1963}.  The finished set-up is shown in the bottom panel of Fig. \ref{Fig:sanae_jpg_pic}.\\

%\textcolor{red}{The value of this resistor does, however, depend strongly on the nature of the pre-amplifier used: The new SANAE set-up uses a simple amplifier, where the (negative) {\it voltage} pulse is amplified, and the 5$\Omega$ resistor minimizes the voltage of noise pulses generated between the walls of the tube and the guard rings. If a charge sensitive pre-amplifier is used, the resistor's value should probably be increased in order to minimize the {\it current} in these noise pulses.} 

When we started examining the 10 BP28 tubes housed at SANAE ({for both the standard 6NM64 and the neutron moderated detector, 4NMD}), it became clear that the tubes all suffered from different levels of degradation (see also the discussion in Sec. \ref{Sec:tube_efficiency} and \ref{Sec:tube_degradation}), with some of the older tubes not producing reliable pulse forms. After testing, 6 tubes were deemed to be in a good condition and we ended up with a 3NM64 (consisting of lead tubes 4,5,6) and a 3NMD (tubes 1,2,3). In this paper we discuss only the 3NM64. Based on the different levels of degradation, tubes 4,5, and 6 have different levels of amplification; the implication thereof discussed in Sec.\ref{Sec:raw_data}.\\

%--------------------------------------------------------------------
\subsection{Software changes}

All the hardware described above uses newly developed software, similar to what is implemented for our mini-NM programme and described in detail by \citet{Straussetal2020}. Raw data (see Sec. \ref{Sec:raw_data}) is saved on solid state drives (SSDs) in Antarctica,  while the (uncorrected)  minute-averaged pressure, temperature, and count rate from each tube is sent to a sever in Potchefstroom, South Africa. There, the data is pressure and temperature corrected (see Sec. \ref{Sec:data_correction}). {SANAE raw and corrected data published at \url{https://fskbhe1.puk.ac.za/neutronmonitor/SANAE/}.} Due to the limited bandwidth, not all raw data can be sent back to Potchefstroom in real-time, and only a selection is available to validate the instrument. SSDs are switched at the end of every year during the annual South African relief voyage in December, and then manually copied onto our servers.\\

%--------------------------------------------------------------------
\section{Data processing}

%--------------------------------------------------------------------
\subsection{Raw data and multiplicities}
\label{Sec:raw_data}

\begin{figure}[!t]
\noindent\includegraphics[width=0.49\textwidth]{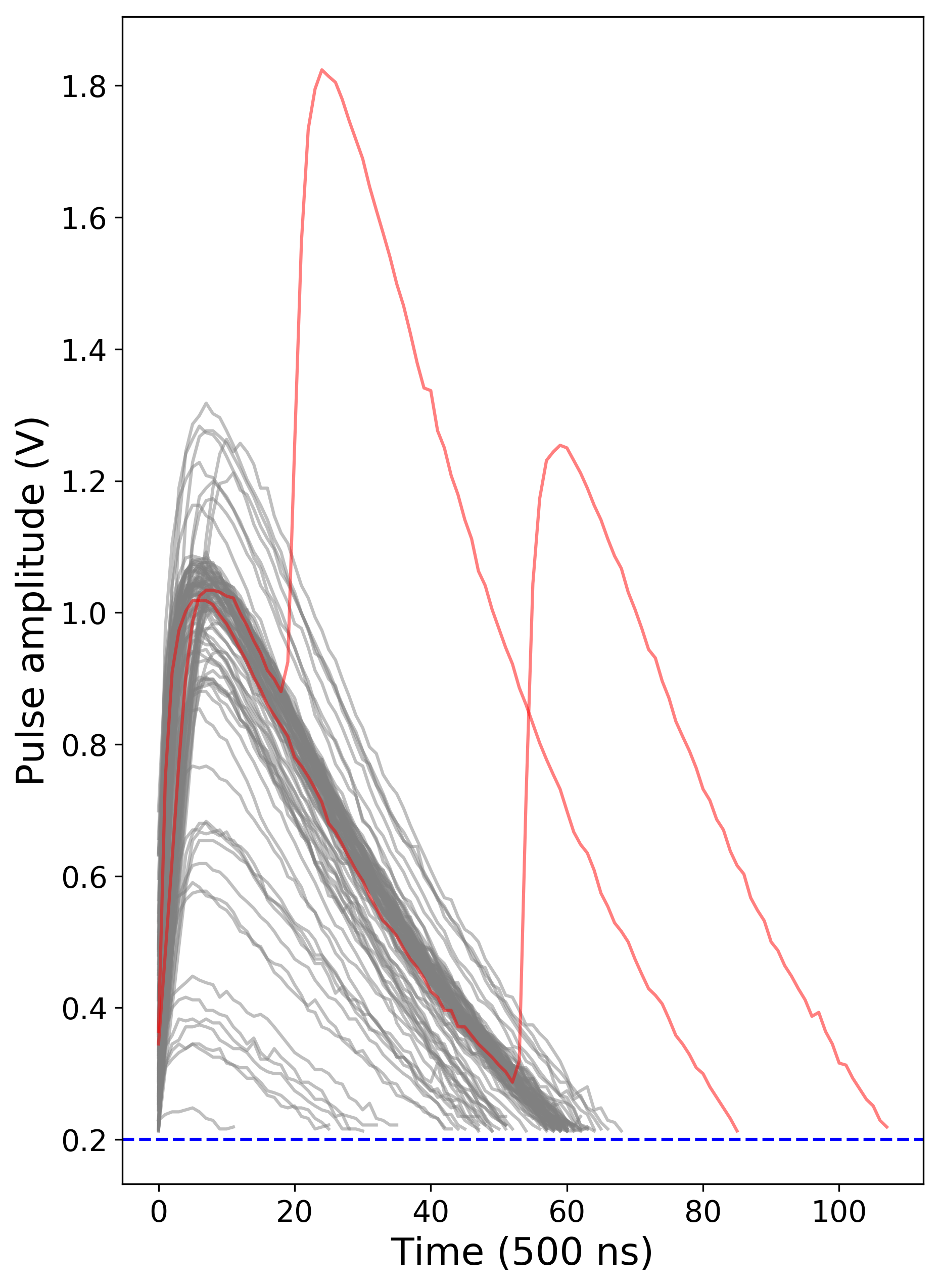}
\caption{{The time profile of 100 pulses, with two overlapping pulses coloured red, is shown. The horizontal dashed line shows the ADC threshold level. Note that the time axis is given in units of 500ns; the native sampling rate of the ADC. }}
\label{Fig:pulses_1}
\end{figure}

\begin{figure}[!t]
\begin{center}
\noindent\includegraphics[width=0.49\textwidth]{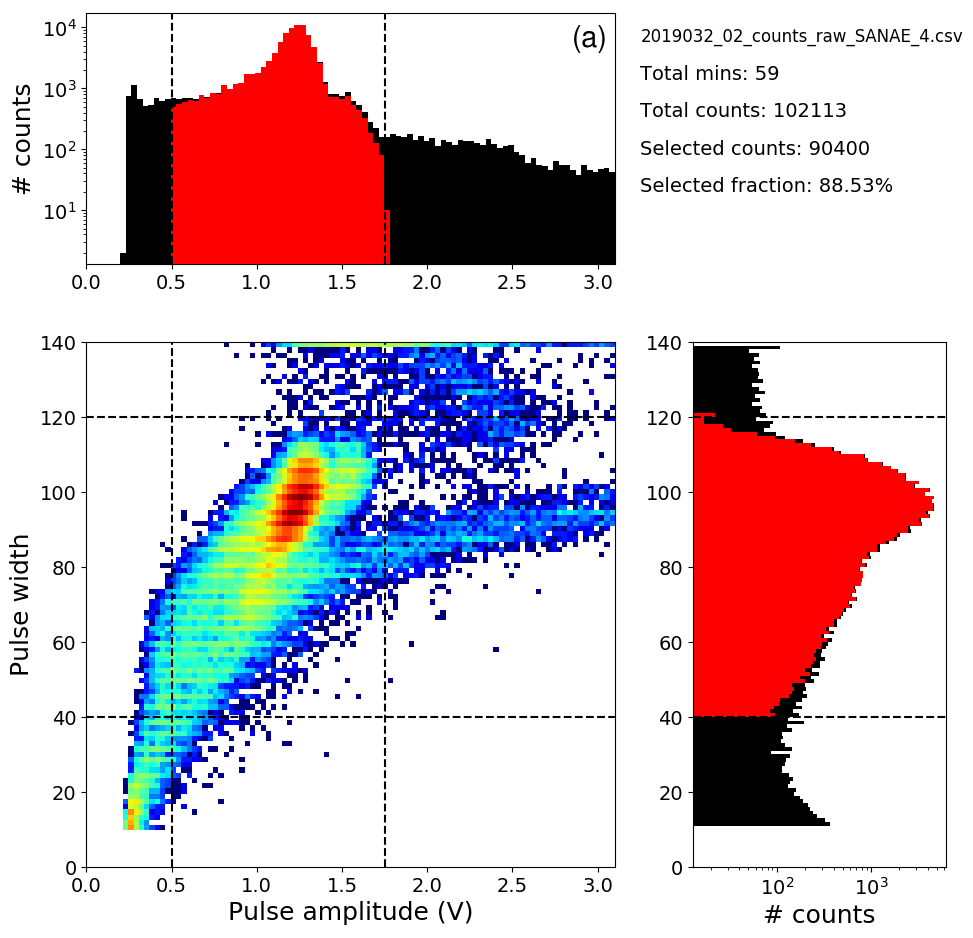}
\noindent\includegraphics[width=0.49\textwidth]{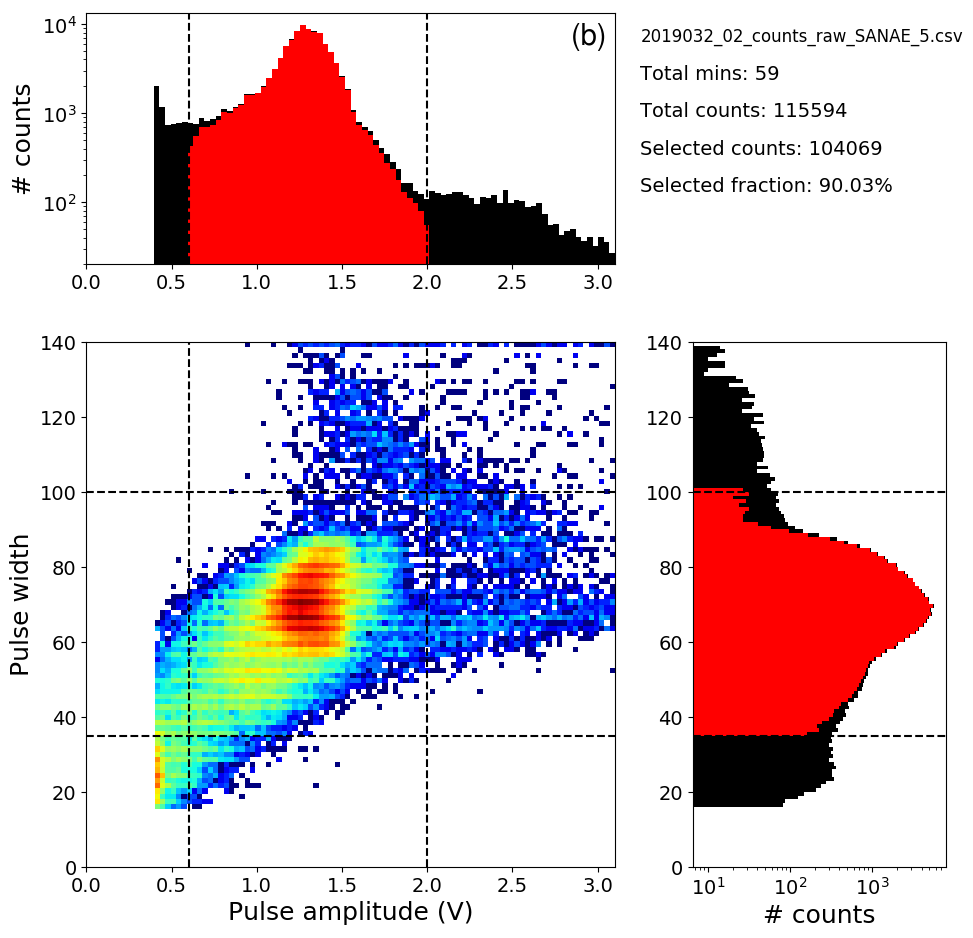}\\
\noindent\includegraphics[width=0.69\textwidth]{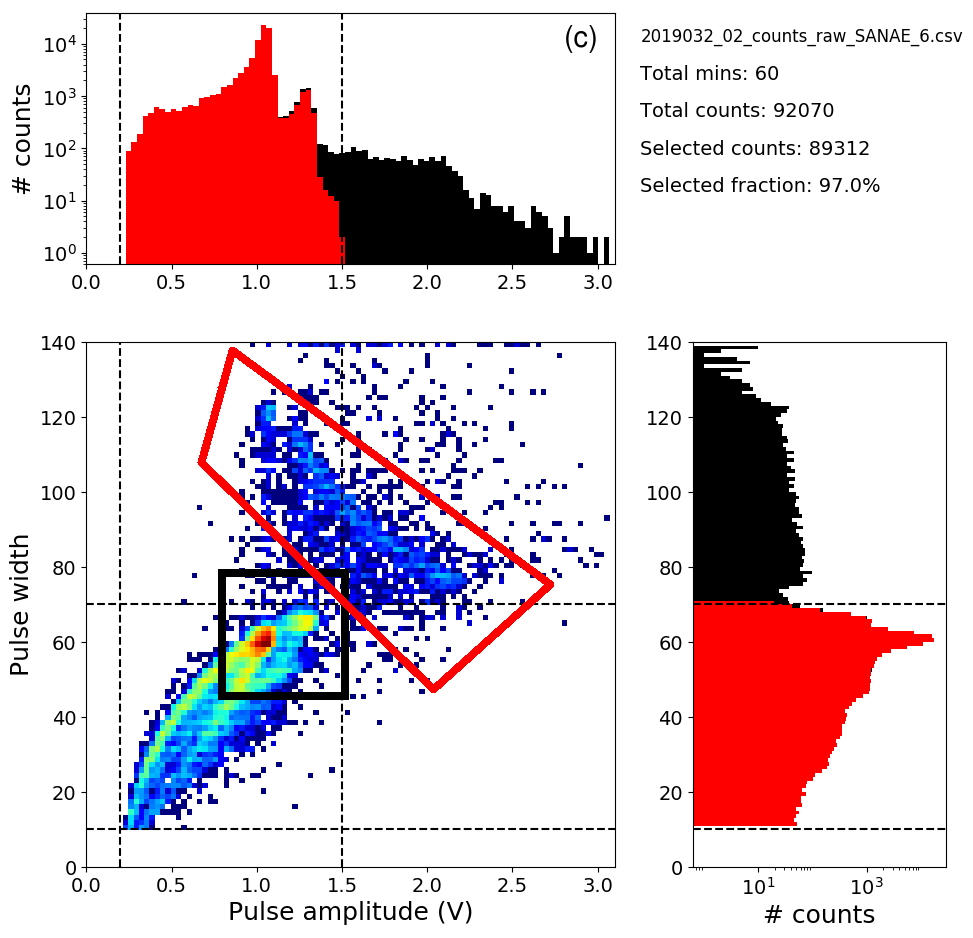}
\caption{{The different panels show the pulse amplitude-width diagrams (number of pulses, in a logarithmic scale, registered with a certain duration (width; in units of 500ns) and pulse amplitude (height) for the 3NM64 tubes presently under consideration (data from tubes 4 (panel a) and 5 (panel b) are shown in the top left and right panels, respectively, and results for tube 6 (panel c) in the bottom panel). The vertical and horizontal dashed lines show examples of how software filters can, in future, be applied.}}
\label{Fig:pulses_2}
\end{center}
\end{figure}

\begin{figure}[!t]
\noindent\includegraphics[width=0.49\textwidth]{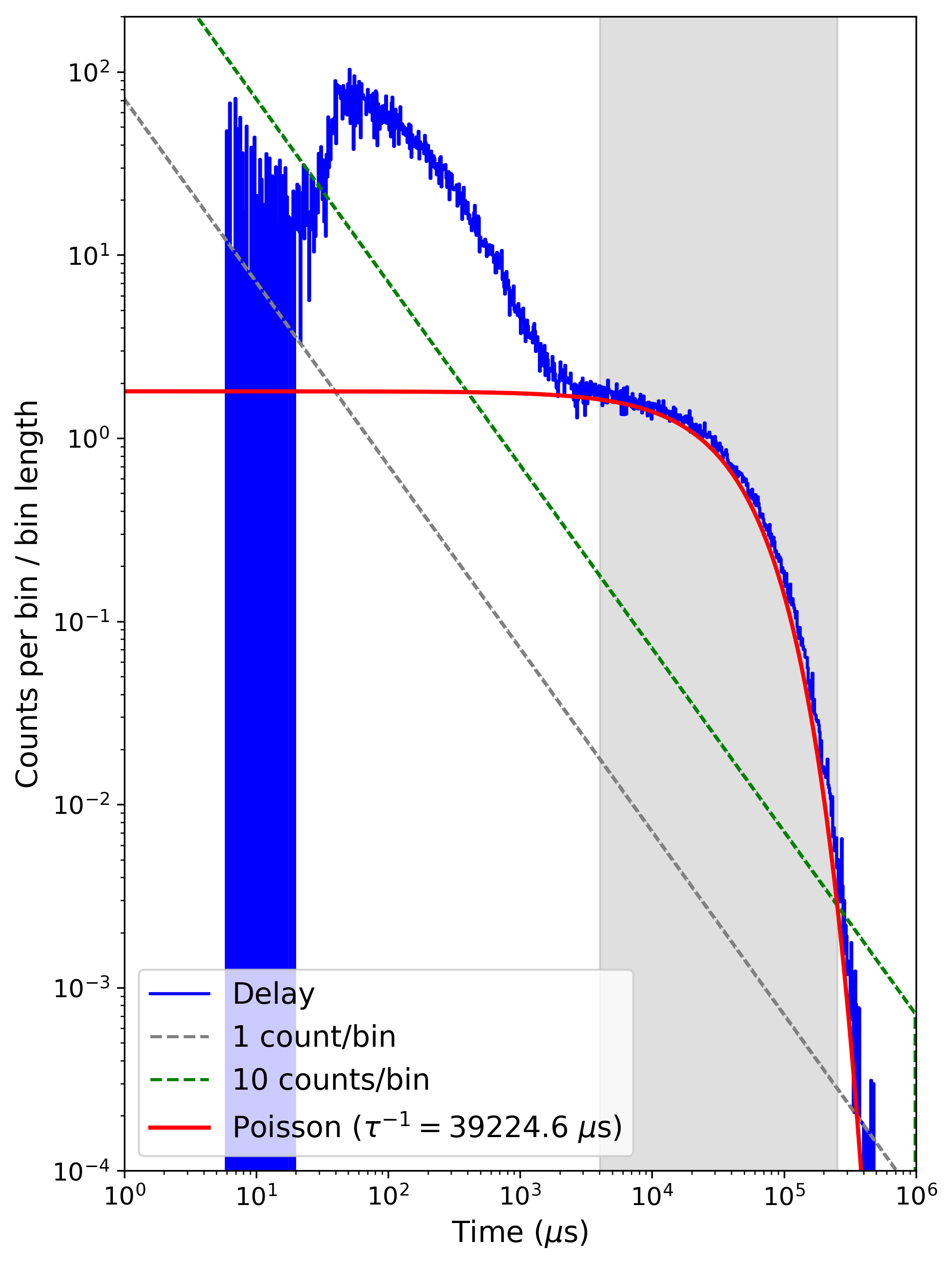}
\noindent\includegraphics[width=0.49\textwidth]{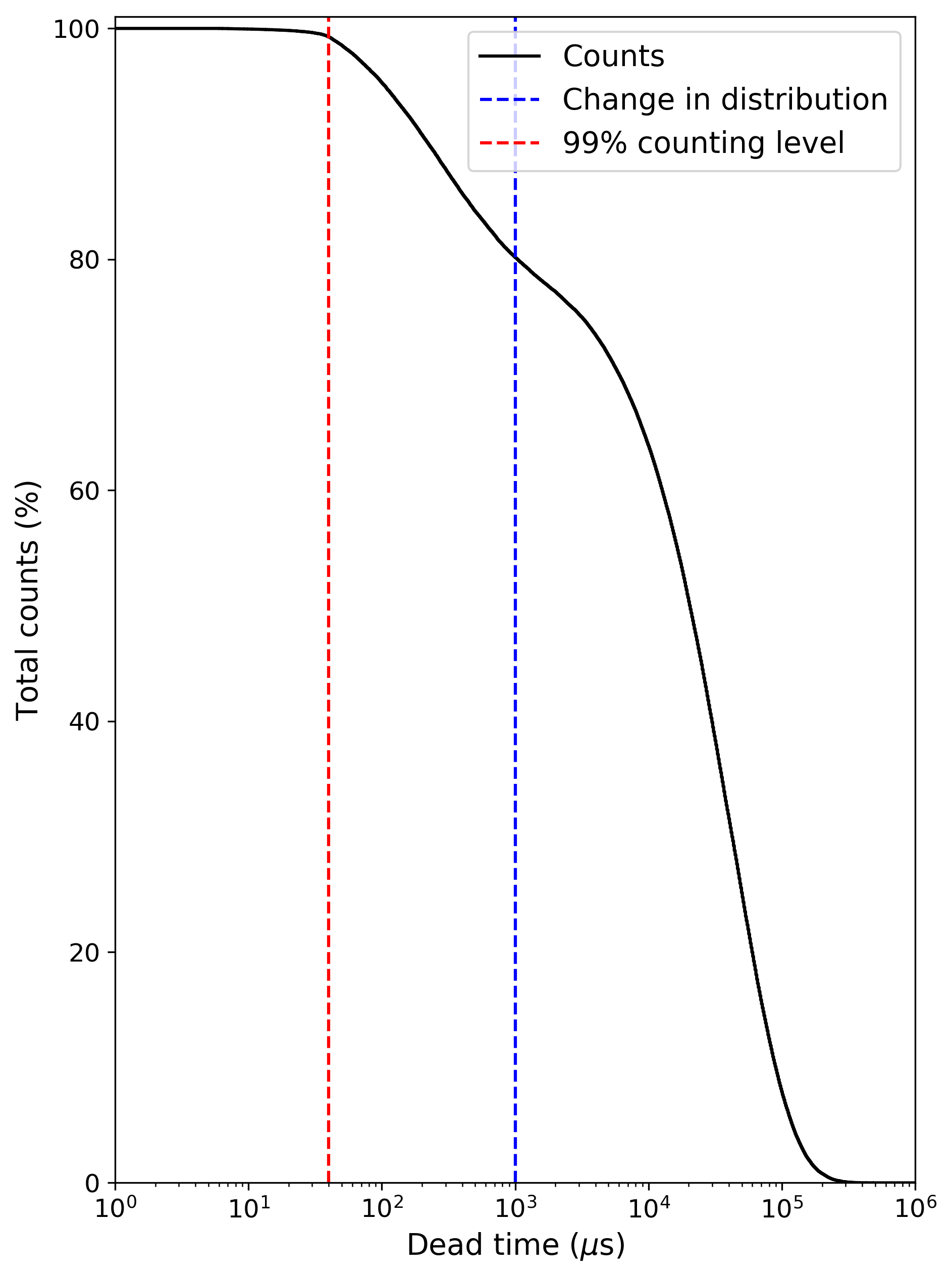}
\caption{The left panel shows the distribution of waiting times (blue histogram) {with the waiting time defined as the time registered between subsequent pulses}. {The solid red curve shows a fitted Poisson distribution.} Green and gray dashed lines show the 1 and 10 count limits. The right panel shows the total integrated count as a function of simulated dead time in the detector.}
\label{Fig:multiplicities}
\end{figure}

The amplified voltage is continuously sampled by the analogue-to-digital converter (ADC) at a rate of 2MHz (i.e. time interval of 0.5$\mu$s). When this voltage crosses a pre-defined software threshold (currently at 0.2V), the start of the pulse is registered, assigned a relative time-stamp, and these voltage values are saved into the ADC's buffer until the voltage again drops below the threshold. The buffer is then readout to a Raspberry Pi micro-computer, along with the absolute and relative time-stamps. We can therefore {digitize} the shape (i.e. temporal profile) of each pulse registered. As far as we are aware, this is currently the only large-scale NM that has this ability. Fig. \ref{Fig:pulses_1} shows 100 of these pulses (in gray) and the ADC threshold level (blue dashed line). The red curves show two occurrences of the interesting scenario where pulses overlap. These events are, however, rare for the current counting rate ($<1\%$). We can also apply further software filters on the digitized pulses when the ADC's buffer is read out; {something that is not implemented at the moment}.\\

{Shown in Fig. \ref{Fig:pulses_2} are pulse amplitude-width diagrams constructed from the individual pulses for each tube. The pulse width (length or duration) is the time {(here, in units of 500ns; the native sampling rate of the ADC)} from which a registered pulse first crosses the ADC threshold to where the signal again decreasing below this threshold, while the pulse amplitude is the maximum voltage of each pulse. The amplifying stage can only amplify pulses up to 3.3V and therefore the amplification and shaping stages are chosen so that the pulses from cosmic ray interactions and detector noise (low amplitude, low width pulses) can be distinguished \citep[see the general discussion in e.g.][for a discussion on how these two distributions can be separated]{knoll12010}}. In order to accomplish this, we had to adjust the amplification of each tube; newer tubes have lower amplification. {This leads to different pulse width-height diagrams for each tube.} These 2D distribution are also projected onto both the pulse amplitude and width planes, shown as the black distributions.  Currently we do not apply any additional data selections on the distributions and simply count all the pulses shown here. However, in future, we can apply more complex data selections to better constrain instrumental noise. As an example, we show a simple rectangular selection (indicated by the dashed lines), with data in this selection shown as the red histograms.\\

{In the bottom panel (c)} we also indicate two effects: {The black rectangle emphasizes a portion of the distribution that indicates two peaks produced by neutron capture in the BF$_3$ tube, which we attribute to the two reactions in Eq. \ref{Eq:reactions}, with the $^{7}\mathrm{Li}^{*}$ reaction producing a peak with a slightly lower amplitude.} The red rectangle shows a selection of pulse which we believe is the population of overlapping pulses (the red pulse in the top left panel). We currently do not correct for these scarce over-lapping pulses. However, with even high counting rates, during e.g. ground level enhancements, the effect from these overlapping pulses may become significant and will be investigated further in future. {A recent study of the different pulse shapes, using a similar electronics set-up for a mini-neutron monitor, is presented by \citet{Similia2021}.} \\

The distributions in Fig. \ref{Fig:pulses_2} are from an hour's worth of data. Due to the low bandwidth connection to SANAE, we cannot copy all the raw data, and only a selection of pulses are copy every day from which such a distribution can be reconstructed {(at present only the first pulse registered at the start of every second is sent back in near real-time)}. 
This is our primary check for data quality on the different tubes. 
{The most recent (``live") plot can be found at \url{https://fskbhe1.puk.ac.za/spaceweather/sanae_tube_4.html} for tube 4, and at similar URLs for the other tubes.}\\

Having access to the raw data, we can also calculated the waiting-time (or delay) distribution of the pulses. This is done by calculating the delay between the onset of subsequent pulses, where $j$ labels the pulse number, as $\Delta t_j = \,^{\mathrm{o}}t_{j + 1} - \,^{\mathrm{o}}t_j$ {where the superscript $\,^\mathrm{o}$ refers to the onset of each pulse}. We use the (more accurate) relative time-stamp of the ADC as we are interested in accurate time differences, and not in the absolute timing of the pulses. The left panel of Fig. \ref{Fig:multiplicities} shows this distribution as a blue histogram for an hour worth of data from tube 4 and is similar to the previous results of \citet{Krugeretal2017}. The gray and green dashed lines show the 1 and 10 count limits. 
The system implemented continuously samples at 2MHz and does not have a longer hardware-define dead-time; the dead-time is determined by the length of the pulses, and, as shown before, different pulses can overlap and be counted as a single detection. 
The blue histogram in the left panel cuts off below $\sim 50 - 100 \mu$s, not due to any dead-time effects, but rather because the average pulse length is reached. \\

Two distinct distributions are visible in the waiting-time distribution: The second distribution, with the larger waiting times, is the population of low-multiplicity neutrons produced both in the atmosphere, and neutrons produced in the lead producer from low energy protons \citep{bieberetal2004,balabin2011,mangeard2016,saiz2019}. The first distribution, with shorter waiting times, are high-multiplicity evaporated neutrons produced in the lead producer during nuclear interactions with high-energy incident protons. Our current system can calculate the multiplicity spectrum to much shorter time delays as compared to contemporary instruments \citep[e.g.][]{ruffoloetal2016}. The low-multiplicity (large delays between pulses) distribution can be well described by a Poisson distribution

\begin{equation}
    \mathcal{D}(\Delta t) = \mathcal{D}_0  \exp \left( \frac{\Delta t - \, ^\mathrm{o}\Delta t}{\tau} \right),
\end{equation}

{with $\Delta t$ the binned waiting times (time between subsequent pulses) and $\,^\mathrm{o}\Delta t$ a constant. This distribution is} shown as the solid red line, and the resulting $\tau$ indicated in the legend. The gray shaded area indicate the data used to constrain the fitted the distribution. \\

In the right panel of Fig. \ref{Fig:multiplicities}, we calculate, using measurements presented in the left panel, the total number of counts measured by this detector as a function of simulated dead time of the detector, $\Delta T$,

\begin{equation}
   \mathcal{C}(\Delta T) = \int_{\Delta T}^{\infty} \mathcal{D}(\Delta t) d (\Delta T).
\end{equation}

The blue vertical line at $\sim 1$ms, indicates the time scale separating the two multiplicity distributions discussed earlier, consistent with the estimate of $\sim 4$ms of \citet{bieberetal2004}. We estimate that at least 20\% of the neutrons detected in the SANAE NM is produced in the lead producer via high-multiplicity evaporated neutrons. This is, of course, expected to be dependent on the incident particle spectrum, and will therefore be time dependent, and dependent on the cut-off rigidity of any NM. The vertical red line, at 40$\mu$s shows the minimum dead time in the detector needed to count 99\% of incident particles. {A typical NM64 set-up has a dead time $\sim 20\mu$s depending on the electronics used \citep[see the discussion by e.g.][]{bieberetal2004}.}.\\
 
%--------------------------------------------------------------------
\subsection{Processed data}
\label{Sec:data_correction}

\begin{figure}[!t]
\noindent\includegraphics[width=0.99\textwidth]{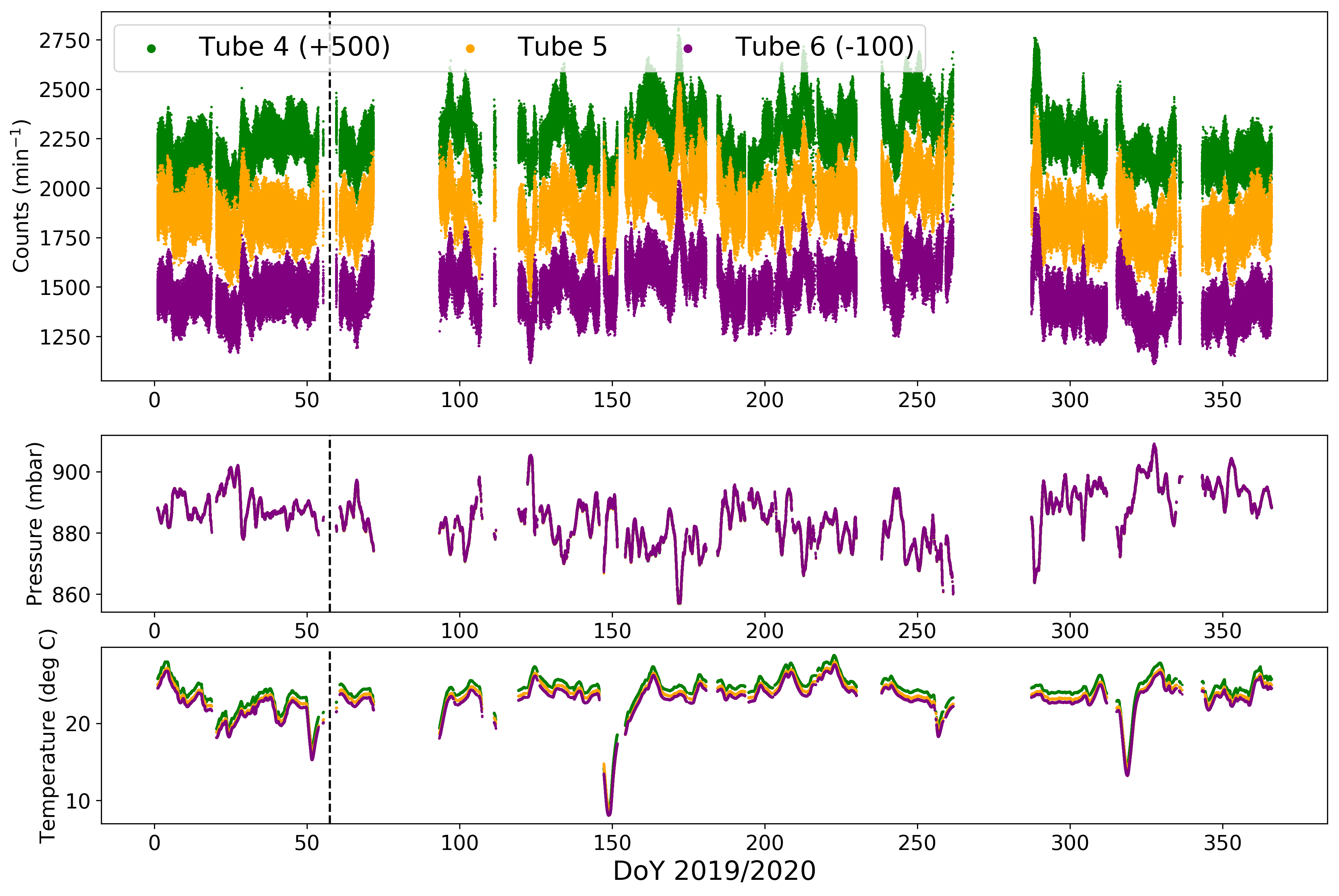}
\caption{The top panel shows the uncorrected count rate from the three working NM tubes, the middle panel the pressure from each tube, and the bottom panel the temperature. {Both the temperature and pressure is measured by a temperature probe inserted next to each BF$_3$ tube.} The data extends from 2019 DOY 60 to 2020 DOY 55. All measurements have a minute cadence.}
\label{Fig:raw_data}
\end{figure}

The main space weather data product from the SANAE NM is the minute-averaged count rate, and we use these measurements from March 2019 to February 2020 in this paper to perform the data corrections {described in this section}. Fig. \ref{Fig:raw_data} shows these measurements for times when all three tubes provided data. The top panel of the figure shows the count rates, the middle panel the pressures, and the bottom panel the temperatures. The data gaps correspond mostly to times when the instrument was off-line for hardware testing and software upgrades. As expected, the pressure values from the different tubes are comparable, and reflect the accuracy of the cost-effective BME-280 pressure sensors used here \citep[see again the description in][]{Straussetal2020} \\

The temperatures of the different tubes are, however, significantly different and there is a clear temperature gradient in the monitor with tube 6 being consistently colder. 
Also note that the temperature of the monitor does change significantly throughout a year, and a temperature correction will have to be performed as discussed later. 
Although the base housing the NM is heated, block C is actively used as a storeroom for the base's vehicles, and the outside roller-doors opened occasionally. 
Also note that during extreme storms (e.g. 2019 DOY 150), the temperature inside block C can drop significantly.\\

%--------------------------------------------------------------------
\subsection{Pressure correction}

\begin{figure}[!t]
\noindent\includegraphics[width=0.99\textwidth]{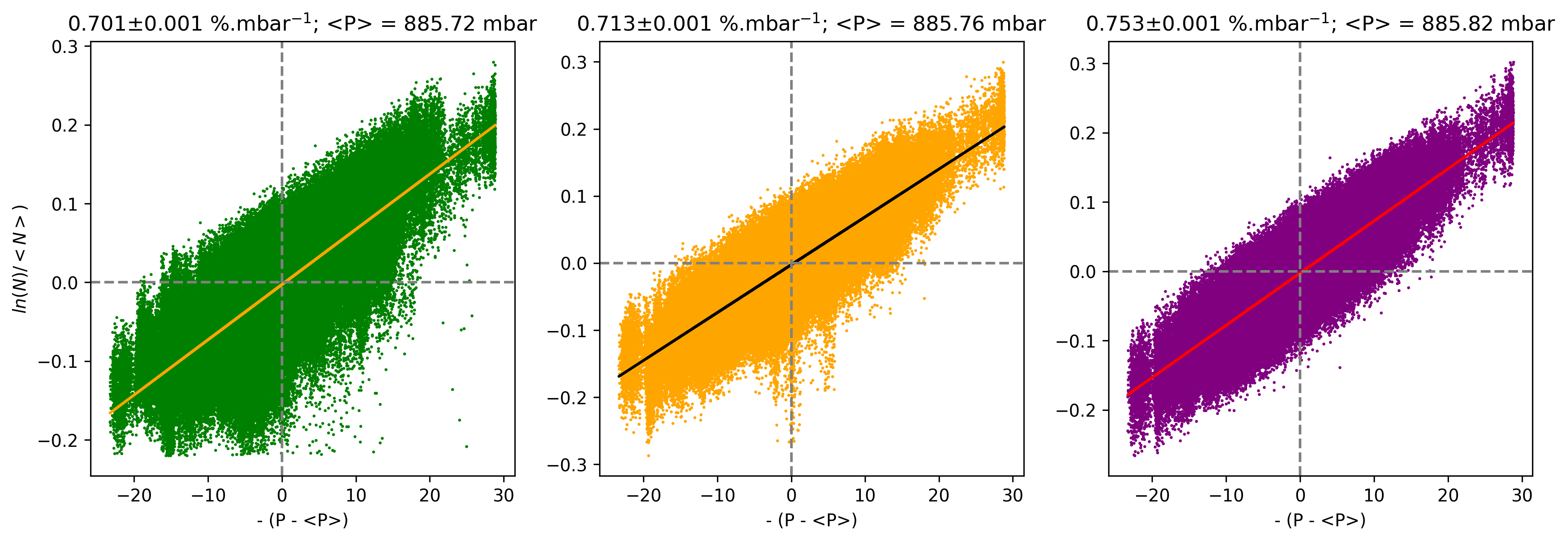}
\noindent\includegraphics[width=0.99\textwidth]{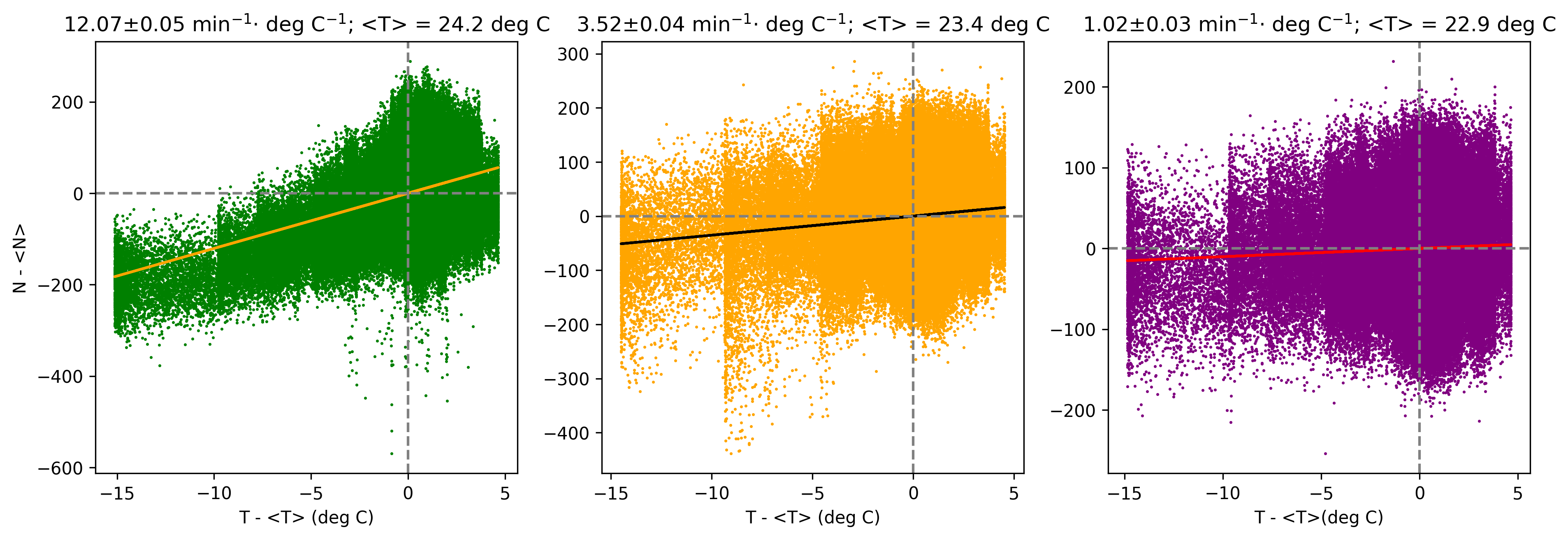}
\caption{The top rows show the pressure correction calculations, and the bottom panel the temperature corrections. The left panels are for tube 4, the middle panels for tube 5, and the right panel for tube 6.}
\label{Fig:pressure_temp_correction}
\end{figure}

Correction for atmospheric changes is the primary correction on the NM count rate \citep{carmichael1968,Paschalisetal2013}. The count rate of each tube $i$ varies according to the relationship,

\begin{equation}
    N_i = \langle N_i \rangle \exp \left(- \beta_i \Delta P_i \right),
\end{equation}

where $\Delta P_i = P_i - \langle P_i \rangle$ is the pressure change, $\langle P_i \rangle$ is a long-term average pressure value, and $\beta_i$ is the {\it barometric coefficient} for each tube, labelled by $i$. To calculate the barometric coefficient of each tube, we plot $\ln (N_i / \langle N_i \rangle)$ as a function of $- \Delta P_i $, and perform a linear regression, with the barometric coefficient being the gradient thereof. This is shown in the top panel of Fig. \ref{Fig:pressure_temp_correction} for the different tubes; the left panels are for tube 4, the middle panels for tube 5, and the right panel for tube 6. The data points show minute averages, while the linear regression are indicated by the straight lines. The calculated barometric coefficients, with their associated errors are indicated in the title of each figure. There is a statistically significant difference in the barometric coefficient of each tube which is presumably related to slightly different responses to the incident particles.\\

Once these coefficients are known, we apply the pressure correction by calculating

\begin{equation}
    N_i' = N_i \exp \left( \beta_i \Delta P_i \right),
\end{equation}

where $N_i'$ is the pressure corrected count rate of each tube. At present we use only the internal pressure of the base and do not correct for possible pressure changes due to changing environmental conditions \citep[see e.g.][for a discussion of such effects]{malanmoraal2002}. \\

%--------------------------------------------------------------------
\subsection{Temperature correction}

After applying the pressure correction, we test for any possible temperature dependence in the count rate. The bottom panel of Fig. \ref{Fig:pressure_temp_correction} shows $N_i'/ \langle N_i' \rangle$ as a function of the temperature change $\Delta T_i = T_i - \langle T_i  \rangle$ for the different tubes. The correction for the temperature changes can be implemented as

\begin{equation}
    ^T N_i' = N_i' - {^T}\beta_i \Delta T_i,
\end{equation}

where ${^T}N_i$ is now the temperature and pressure corrected count rate, and ${^T}\beta_i$ the so-called {\it temperature coefficient} (i.e. the gradient of the linear regression curve). These coefficients are indicated on the graphs for each tube. Again, they are slightly different for each tube, which is again related to internal differences inside the NM \citep{krugeretal2008,evensonetal2005}. Note that, as the temperature of the tubes do not vary significantly on short timescales, we can apply a linear temperature correction and not the normal exponential form \cite[e.g.][]{krugeretal2008}. These forms are, however, consistent as long as ${^T}\beta_i \Delta T_i \ll 1$. \\

%--------------------------------------------------------------------
\subsection{Tube ratios}
\label{sec:tube_ratio}

\begin{figure}[!t]
\noindent\includegraphics[width=0.99\textwidth]{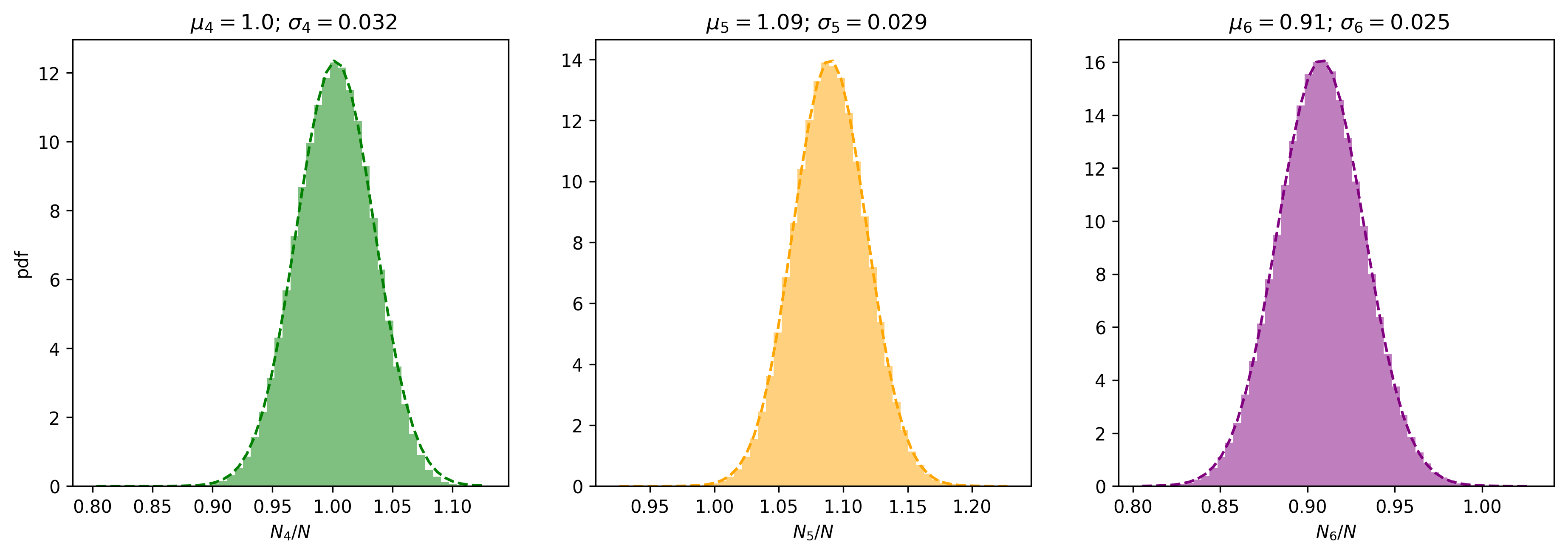}
\noindent\includegraphics[width=0.99\textwidth]{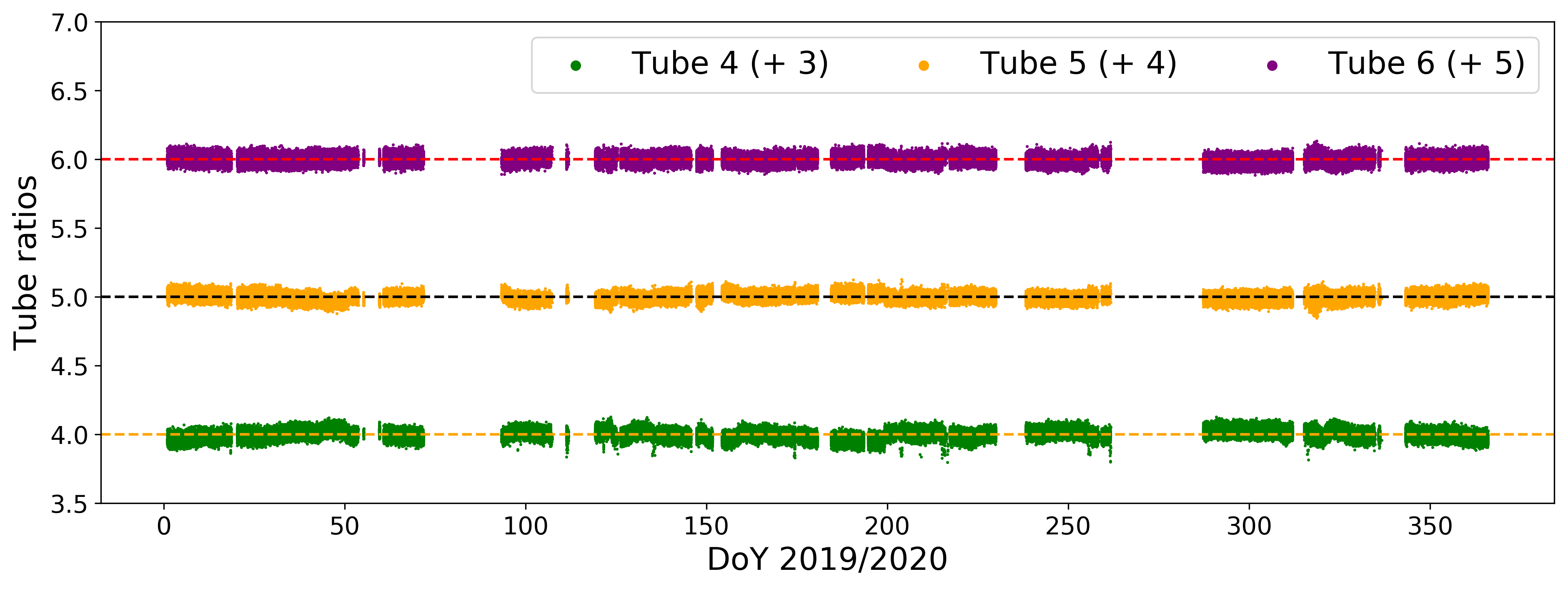}
\caption{The top panel shows binned averages of the calculated tube ratios, for each tube, along with a fitted Gaussian distribution. The bottom panel shows the tube ratio (offset with constant factors) as a function of time.}
\label{Fig:tube_ratio}
\end{figure}

In order to combine the data from different tubes into a single count rate, we make use of so-called {\it tube ratios} \citep{medina2013}. We calculate this quantity for each tube $i$, out of a total of $M$ tubes, as

\begin{equation}
\label{Eq:tube_ratios}
\mu^{-1}_i = \left\langle \frac{1}{{^T}N_i'} \frac{1}{M}  \sum_{i=1}^{M} {^T}N_i' \right\rangle,
\end{equation}

This is the inverse of the {\it average} contribution of each tube to the {\it average} count rate, per tube, of the detector. Usually, in large NMs with identical tubes, we see that the tubes at the edges have a lower count rate than centrally located tubes. This {\it edge effect} is due to the central tubes being surrounded by a larger volume of neutron-producing lead. The tube ratios have the useful relationship

\begin{equation}
\sum_{i=1}^{M} \mu_i = M.
\end{equation}

The top panel of Fig. \ref{Fig:tube_ratio} shows the tube ratios, as calculated for each tube, for the entire dataset presented here. Minute averages are used, binned into histograms, and fitted by a Gaussian distribution, indicating that any variations are due to random fluctuations. Each panel is labelled by the average value of the Gaussian distribution (this is the tube ratio according to Eq. \ref{Eq:tube_ratios}), along  with the variance of the distribution ($\sigma$) given an indication of the error associated with $\mu$.\\

Plotted in the bottom panel is a test quantity

\begin{equation}
 \left\langle \frac{{^T}N_i'}{\mu_i} \left( \frac{1}{M}  \sum_{i=1}^{M} {^T}N_i' \right)^{-1} \right\rangle  = 1.
\end{equation}

These values (related to the contribution of each tube to the total count rate) should, of course, be constant over time, serving as a measure of the validity of the measurements and the stability of the monitor over longer time
scales. {The near real-time SANAE data, including the tube ratios, can be monitored at this page: \url{https://fskbhe1.puk.ac.za/spaceweather/sanae.html}.}

%--------------------------------------------------------------------
\subsection{Tube efficiency}
\label{Sec:tube_efficiency}

\begin{figure}[!t]
\noindent\includegraphics[width=0.99\textwidth]{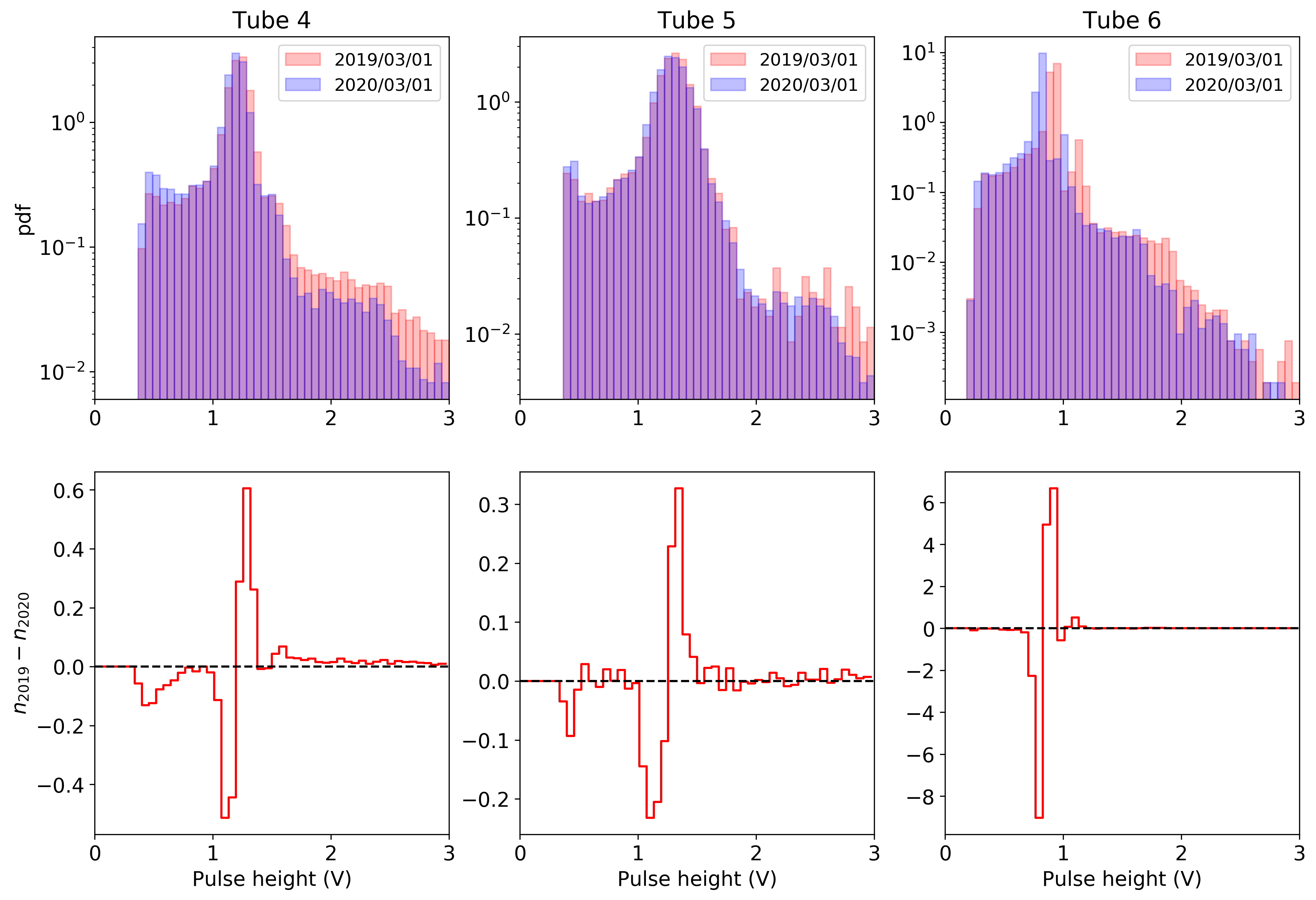}
\caption{The top panels show {normalized} pulse height distributions, for each, tube, taken 12 months apart. The bottom panels show the difference between these distributions.}
\label{Fig:tube_efficiency}
\end{figure}

As already mentioned, our new hardware and software allow us to examine the pulse height distribution on a daily, and even hourly, basis. After running the new system for 12 months, we can also compare, and search for, any possible time dependent degradation of any tubes. In Fig \ref{Fig:tube_efficiency} we show, for each tube, a normalized histogram of the observed pulse height for two days (the first of March) in 2019 and 2020. The 2019 distribution, $n_{\mathrm{2019}}$, is shown in red, and the 2020 distribution, $n_{\mathrm{2020}}$, in blue. Although on a very shorter time scale, these results seem compatible with those presented by \citet{Bieberetal2007} showing tube degradation on a yearly basis, where the peak of the distribution moves to lower voltages, while the distribution becomes wider. As discussed by \citet{Bieberetal2007}, the degradation is probably due to uneven plating of the tube's {central wire}, decreasing its counting efficiency. We present an analysis of several older (and discarded) tubes in \ref{Sec:tube_degradation}. In the bottom panels of the figure, we show $n_{\mathrm{2019}} - n_{\mathrm{2020}}$, illustrating the effect discussed above in more detail. At present, even though the distribution has shifted to lower peak values, all pulses, of cosmic ray origin, still seem to be within the range where they will be counted. As such, we do not yet perform any {\it tube efficiency} calculations to account for any degradation.\\

%--------------------------------------------------------------------
\subsection{Combined count rate}

\begin{figure}[!t]
\noindent\includegraphics[width=0.99\textwidth]{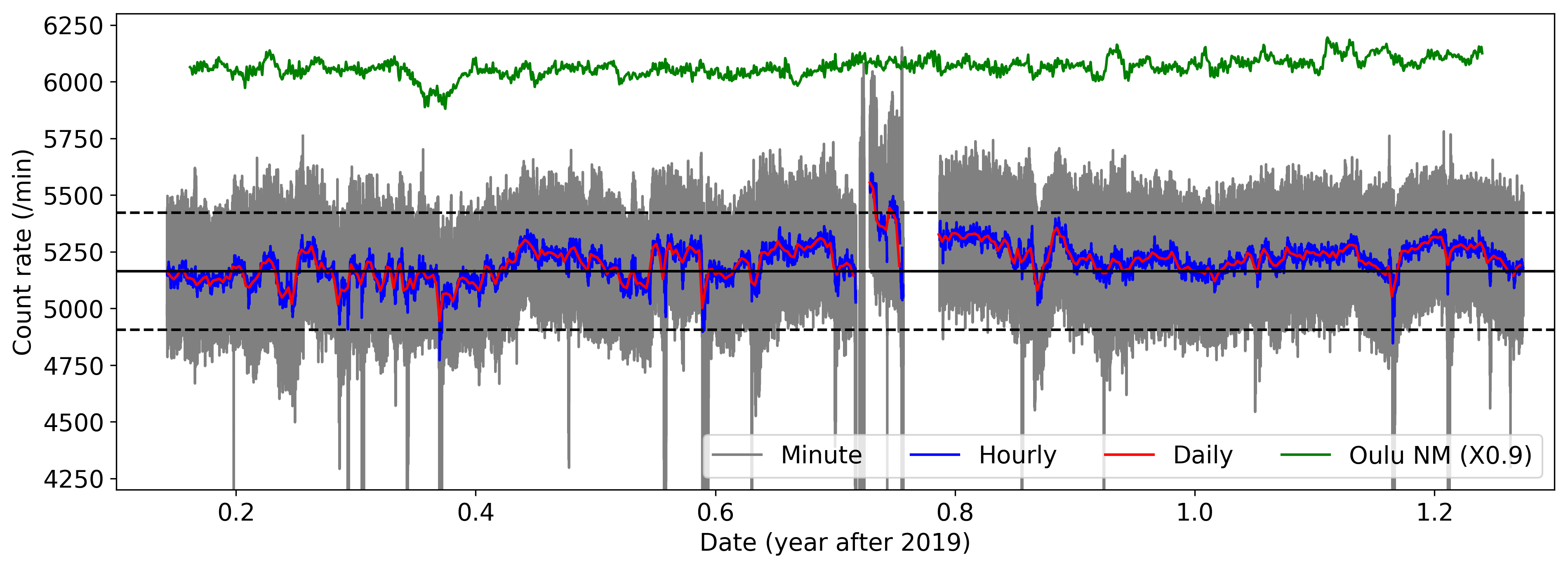}
\noindent\includegraphics[width=0.99\textwidth]{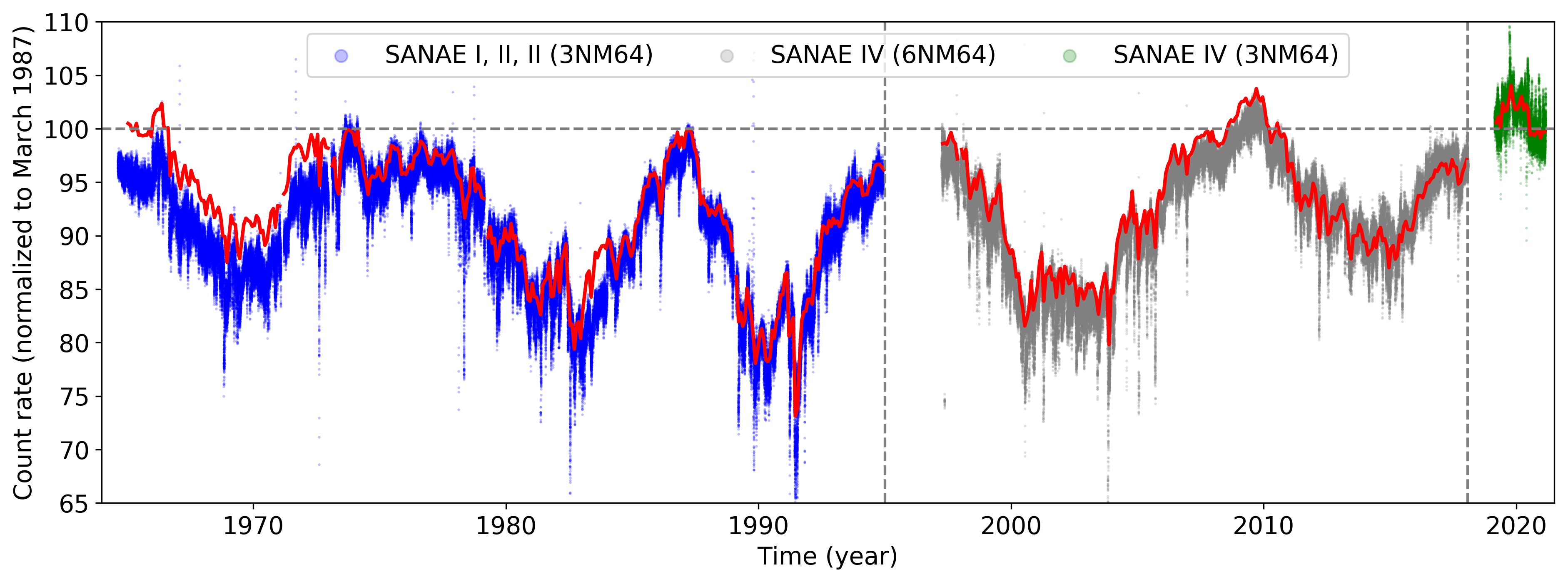}
\caption{The top panel shows the combined SANAE NM count rate from March 2019 to March 2020 in different temporal resolutions, alongside {hourly} data from the Oulu NM. The bottom panel shows the long term SANAE NM observations from 1965 -- present. The symbols are hourly averages and the solid lines are monthly averages. Note the data gaps around 1997 and 2018, corresponding to moving the NM to its current positions, and performing the upgrades presented here, respectively.}
\label{Fig:corrected_data}
\end{figure}

Using the tube ratios, determined in Sec. \ref{sec:tube_ratio}, we can now combine the counts from the different tubes into a single count rate from the monitor. However, we need to account for the fact that not all the tubes will produce valid data at any given time: A tube might be switched off for repairs, there might be spikes from a given tube, or it may be malfunctioning. With $M$ the total number of tubes, and $M'$ the number of tubes active at any given time, we propose the following expression to calculate the total count rate

\begin{equation}
C = \epsilon \left( \frac{M}{M'} \right) \sum_{i = 1}^{M'} \zeta_i \frac{{^T}N_i'}{\mu_i}.
\end{equation}

With the tube ratios correctly calculated, each tube contributes exactly a fraction of $1/M'$ to the total count rate, and the factor $M/M'$ corrects for the possibility that $M'$ might be time dependent. {For a real-time visualization of the number of active SANAE NM tubes, see \url{https://fskbhe1.puk.ac.za/spaceweather/sanae.html}.} We introduce the tube efficiency via the parameter $\zeta_i$, with is currently set to unity for each tube, but could increase/decrease as the degradation of the tubes becomes measurable. The {\it detector efficiency}, $\epsilon$, which is also currently set to unity, can be used, in future, to account for changes in the NM's environment, i.e. changes that will affect the counting rate of all tubes in a similar fashion.\\

The combined SANAE count rate is shown in the top panel of Fig. \ref{Fig:corrected_data} in minute, hourly, and daily resolution as the gray, blue, and red curves. The horizontal lines show the yearly average (solid black line) and possible 5\% variations (dashed black lines). The Oulu NM count rate is shown in green for comparison. During the calibration time, March 2019 -- March 2020, there were very little cosmic ray modulation observed, and this seems to be also reflected in the SANAE observations. There seems, however, to be a small discrepancy in the SANAE NM data around September 2019 (2019.75), which is still under investigations. We currently believe that this might be due to extreme environmental conditions, and this is discussed in more detail in Sec. \ref{Sec:sumary}.\\

%--------------------------------------------------------------------
\subsection{Long-term normalization}

With the recent upgrades and changes to the SANAE NM set-up completed, the current count rate must be normalized to older measurements to create a consistent long-term dataset. In order to do so, we use the Oulu NM as a calibrator: The Oulu NM count rate is normalized to the SANAE dataset (using monthly averages) between January 2009 -- December 2009, where after the new SANAE measurement (monthly averages between March 2019 -- March 2020) is again normalized to the Oulu measurements. Both time periods represent quiet-time solar minimum conditions, with very few temporal variations. The results are shown in the bottom panel of Fig. \ref{Fig:corrected_data}, with the SANAE NM count rate normalized to 100\% in March 1987. The symbols show hourly averages from 1965 to the present, while the solid lines show monthly averages. Note that the dataset is divided into three periods, based on the evolution of SANAE NM, with data gaps around 1997 (when the NM was moved to its current position) and 2018 (when the upgrades presented here were implemented). Note that before 2019 efficiency calculations were only computed on the monthly averages, while the hourly averages were only corrected for pressure changes. This difference is apparent on the figure as a time-dependent difference between the symbols and solid lines, especially diverging towards the start of observations. {These older efficiency calculations were done by hand and, unfortunately, we do not have access to the efficiency calculations pre-2019 and thus cannot correct the minute data retroactively. A major aim of this paper is to also standardize and document the process to remove any such uncertainties in future.}\\

%--------------------------------------------------------------------
\section{Summary and outlook}
\label{Sec:sumary}

In this paper we have described the recent upgrades performed on the SANAE NM in Antarctica. We have retrofitted the older BP28 tubes with newly developed electronics, originally designed for use in our mini-NM programme. These electronics feature, amongst other new developments, a self-contained high-voltage supply, and a new digitizing system allowing us to digitise individual pulses with a 500ns time resolution. This allows us to validate, on an hourly basis, the measured pulses via a pulse width-amplitude plot, and calculate every tube's multiplicity down to the low $\mu$s level. We can also use this to monitor, over longer time-scales, any possible tube degradation. In addition, we have described, in detail, the methodology we applied to correct and combine the data from the different tubes. This is in an effort to standardize the approach used at different NM stations and we will implement these changes to the South African NM network in due course. \\

We are currently looking at possible environmental issues in more detail. We suspect the slight enhancement (up to 5\%) in the SANAE count rate in September 2020 (2020.75), visible in Fig. \ref{Fig:corrected_data} is partially due to a long-lasting, and very severe, storm influencing the detector. These storms at the SANAE IV base usually lead to a large drop in temperature and pressure, and we will need to make sure our current approach is valid for such extreme conditions. In addition to changing environmental conditions, the large wind gusts associated with the Antarctic storms can lead to significant mechanical vibration in the base, and this might contribute to mechanical noise in the detector electronics. We will report on any such issues in future publications. As a preliminary quality flag the number of tubes registering counts at a specific time can be used as excessive vibrations usually lead to saturation of at least one tube's ADC. The near real-time calculated tube ratios are also a good indication of the data quality for detailed studies. \\

The upgrading process described here can be used on any NM using the older BP28 tubes, as well as the even-older IGY (International Geophysical Year) tubes. We have started testing and installing the new hardware on the Potchefstroom IGY NM, and have not encountered any major issues. Based on our experience, we are therefore confident that the upgrades described in this paper can be used to rejuvenate the current ailing NM network, and even bring older, decommissioned, NMs back on-line. This can help with efforts to expand the current world-wide network of NMs \citep[see e.g.][]{Mishev_Usoskin_2020}.\\

A 1NMD (neutron moderated detector, i.e. a NM without the lead producer, also referred to as a ``bare" NM) was installed in at SANAE in 1971 and upgraded to a 4NMD in 1973. Currently, this detector is operated as a 3NMD, upgraded with the same electronics as discussed in this paper in early 2020. We are currently gathering calibration data and expect measurement to become publicly available later in 2021. Live data from the different tubes are, however, already available at e.g. \url{https://fskbhe1.puk.ac.za/spaceweather/sanae_tube_1.html} for tube 1, with similar links to tubes 2 and 3.\\

%--------------------------------------------------------------------

% >>>>	A statement that indicates to the reader where the data
% 	supporting the conclusions can be obtained (for example, in the
% 	references, tables, supporting information, and other databases).
%
% 	All funding sources related to this work from all authors
%
% 	Any real or perceived financial conflicts of interests for any
%	author
%
% 	Other affiliations for any author that may be perceived as
% 	having a conflict of interest with respect to the results of this
% 	paper.
%
%
% It is also the appropriate place to thank colleagues and other contributors.
% AGU does not normally allow dedications.

\section*{Acknowledgement}

Oulu data obtained from \url{http://cosmicrays.oulu.fi}, courtesy of the Sodankyl{\"a} Geophysical Observatory. Figures prepared with Matplotlib \citet{Matplotlib-2007} and certain calculations done with NumPy \citep{harrisetal2020}. This work is based on the research supported in part by the National Research Foundation of South Africa (NRF grant numbers: 119424, 120847, and 120345). Support from the NRF's South African National Antarctic Programme (SANAP; grant number 110741) is especially acknowledged. Opinions expressed and conclusions arrived at are those of the authors and are not necessarily to be attributed to the NRF.

%-------------------------------------------------------------
\appendix

%-------------------------------------------------------------
\section{Tube degradation}
\label{Sec:tube_degradation}

\begin{figure}[!t]
\begin{center}
\noindent\includegraphics[width=0.32\textwidth]{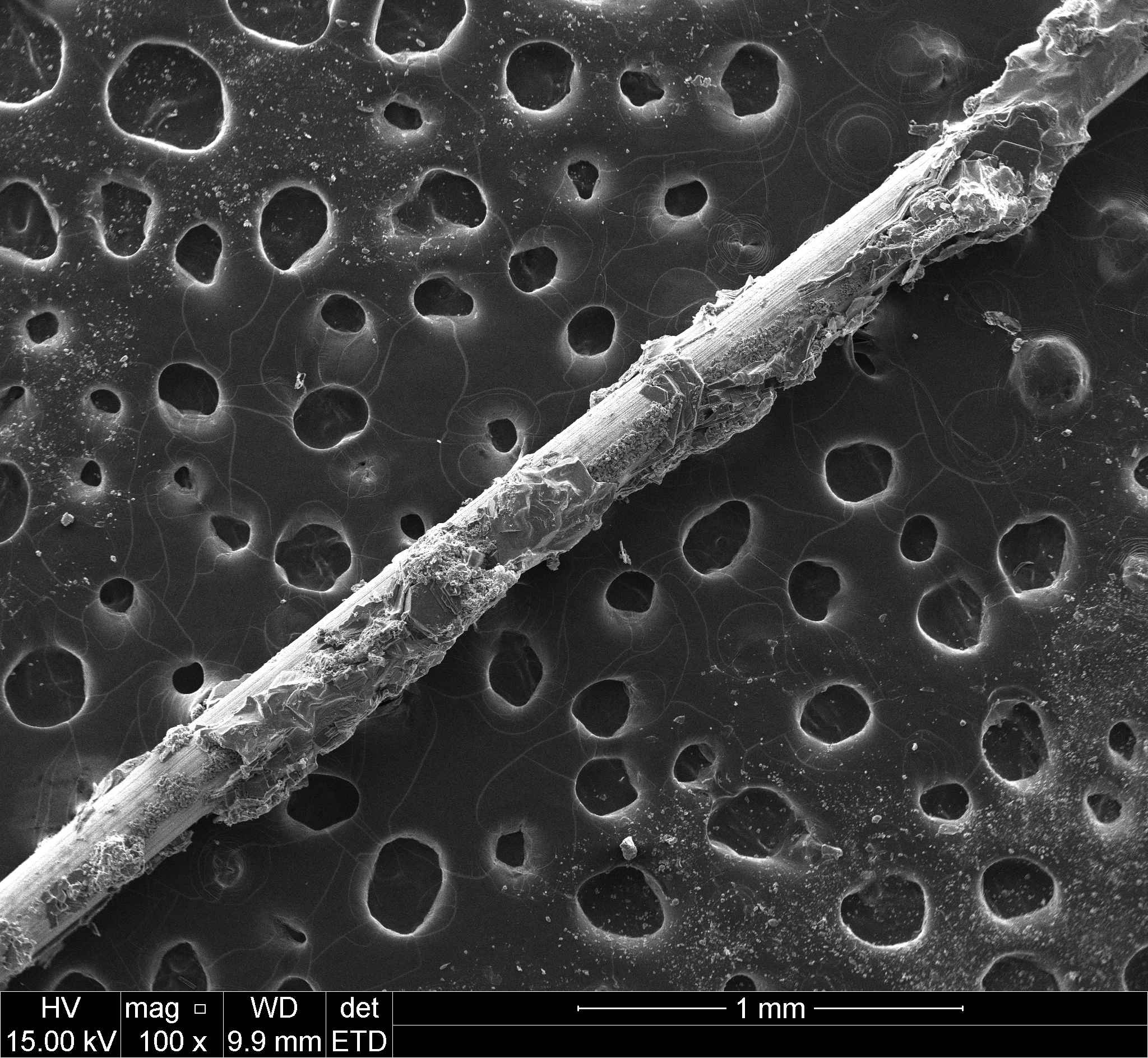}
\noindent\includegraphics[width=0.32\textwidth]{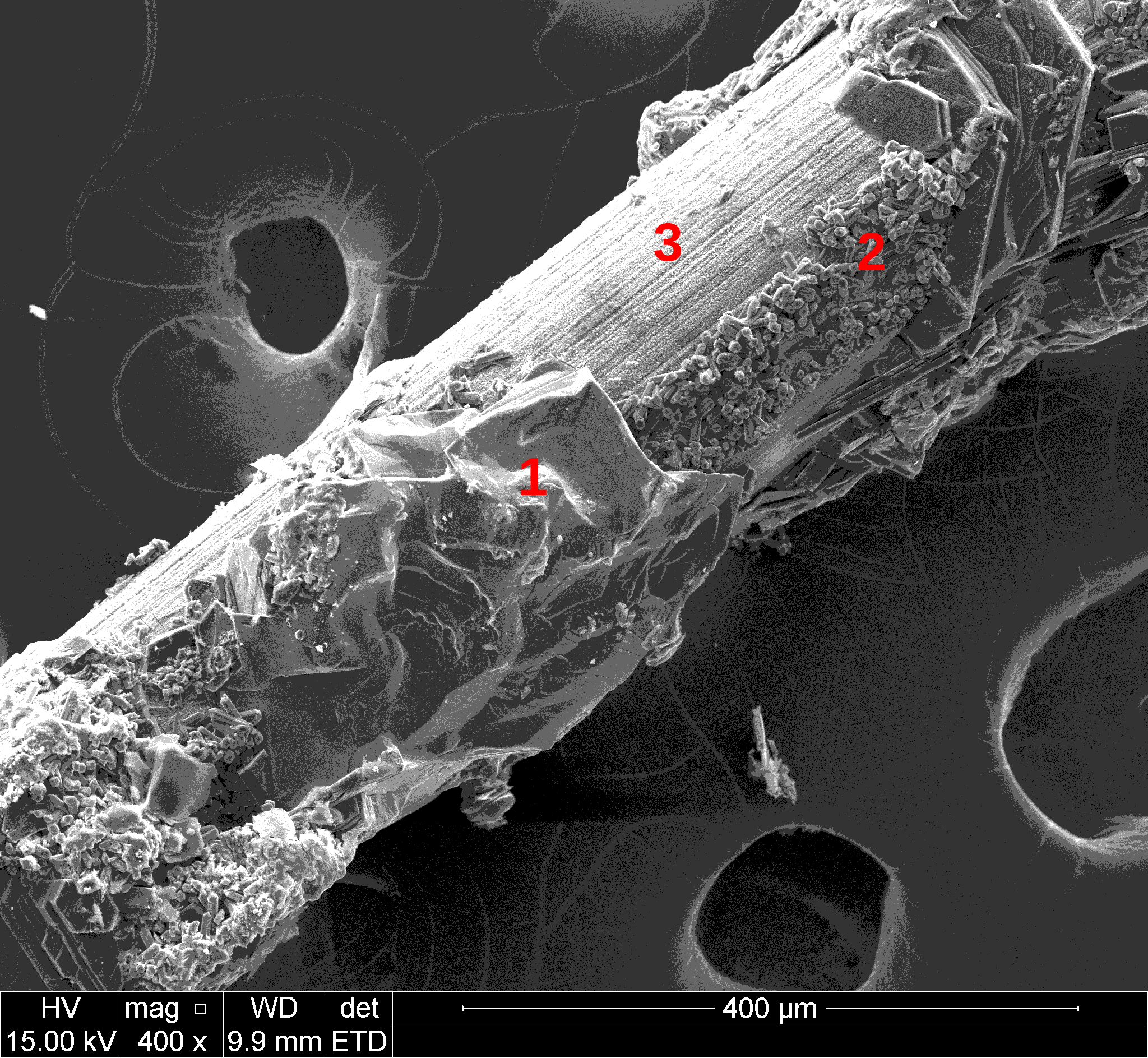}
\noindent\includegraphics[width=0.32\textwidth]{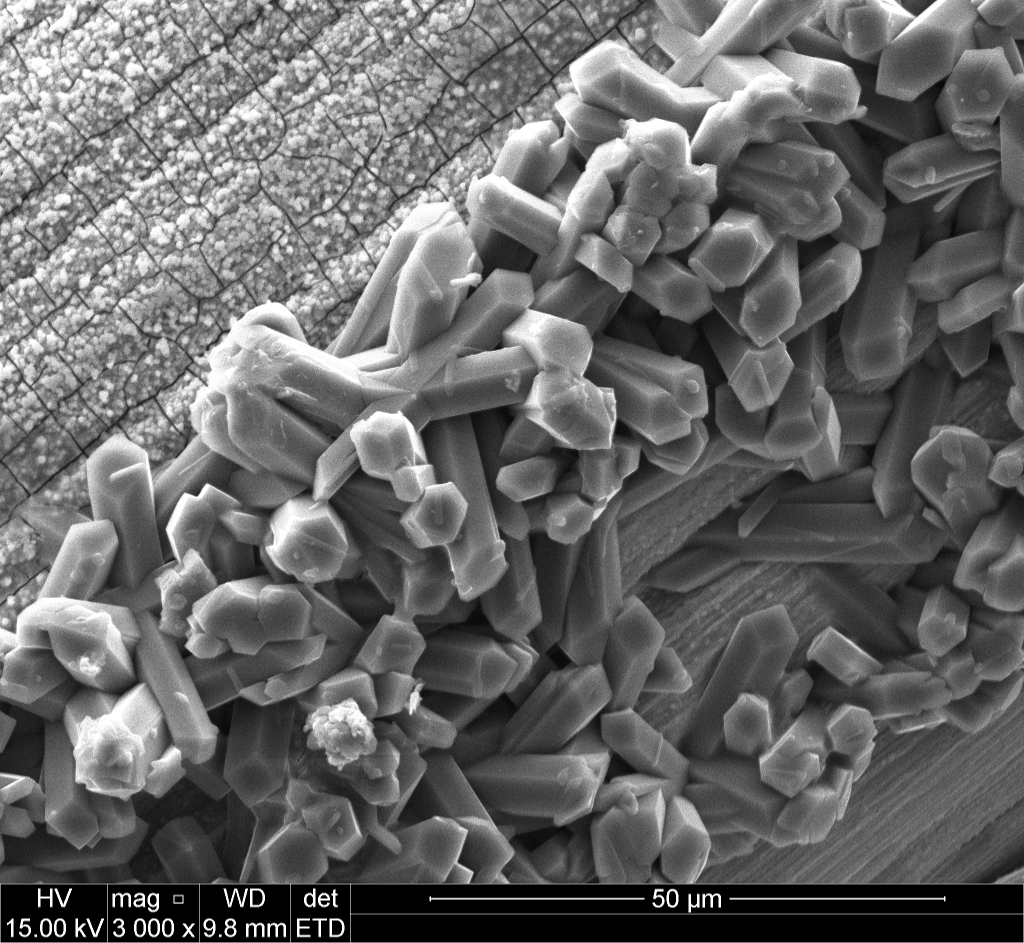}
\caption{An example of the crystal structures found on several {central wires}.}
\label{Fig:tube_microscope}
\end{center}
\end{figure}

As mentioned in the main text, during the upgrading of the SANAE NM, several of the older NM tubes were discarded as they either provided no, or very weak pulses. The tubes were transported back to Potchefstroom and investigated in more detail. Of course, there are several reason for a tube to malfunction and completely stop registering pulse, including physical damage to the {central wire} (the integrity of the {central wires} can be confirmed by measuring the impedance of the tube) or a broken gas seal, leading to escape of the BF$_3$ gas. However, based on the results of Sec. \ref{Sec:tube_efficiency}, we are also interested in finding possible reasons for the steady decline of the count rate of tube with apparently no physical damage. {The central wire} of the broken/inefficient tubes were removed and examined under a scanning electron microscope for any discrepancies. We noted several that the {central wires} were non-uniformly encrusted with crystal structures. An example is shown in Fig. \ref{Fig:tube_microscope} under different levels of magnification. In the middle panel, we identified three possible different crystal structures; a very coarse structure (labelled 1), a finer structure (labelled 2), and a thin layer of plating (labelled 3).

\begin{figure}[!t]
\begin{center}
\noindent\includegraphics[width=0.38\textwidth]{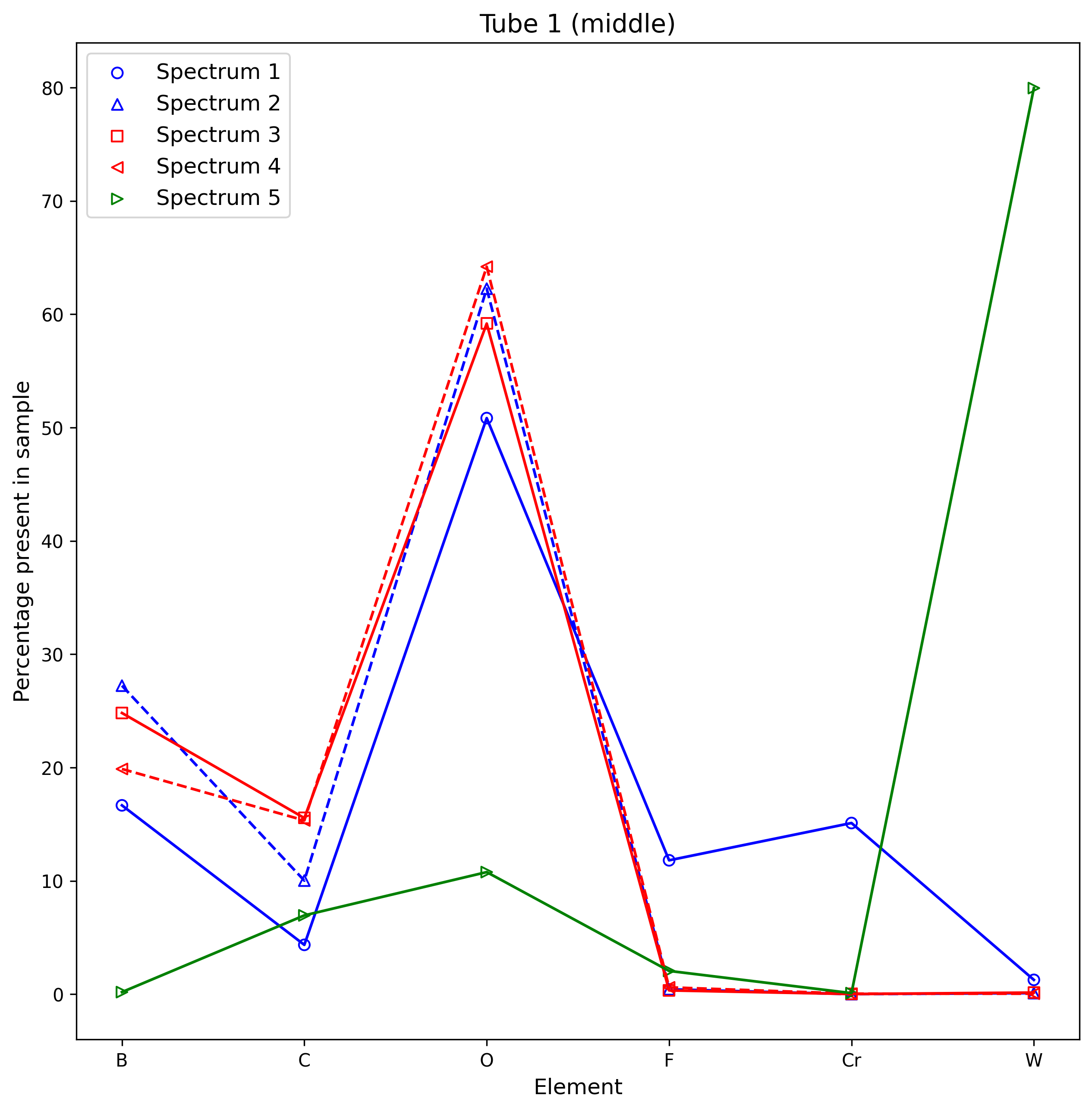}
\noindent\includegraphics[width=0.4\textwidth]{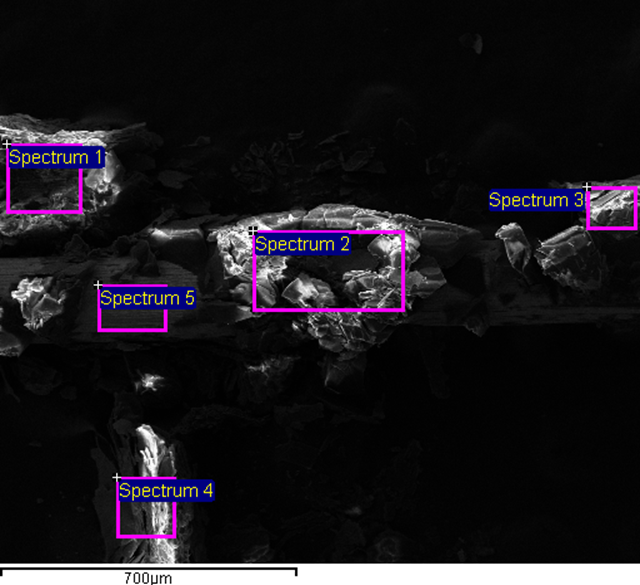}\\
\noindent\includegraphics[width=0.38\textwidth]{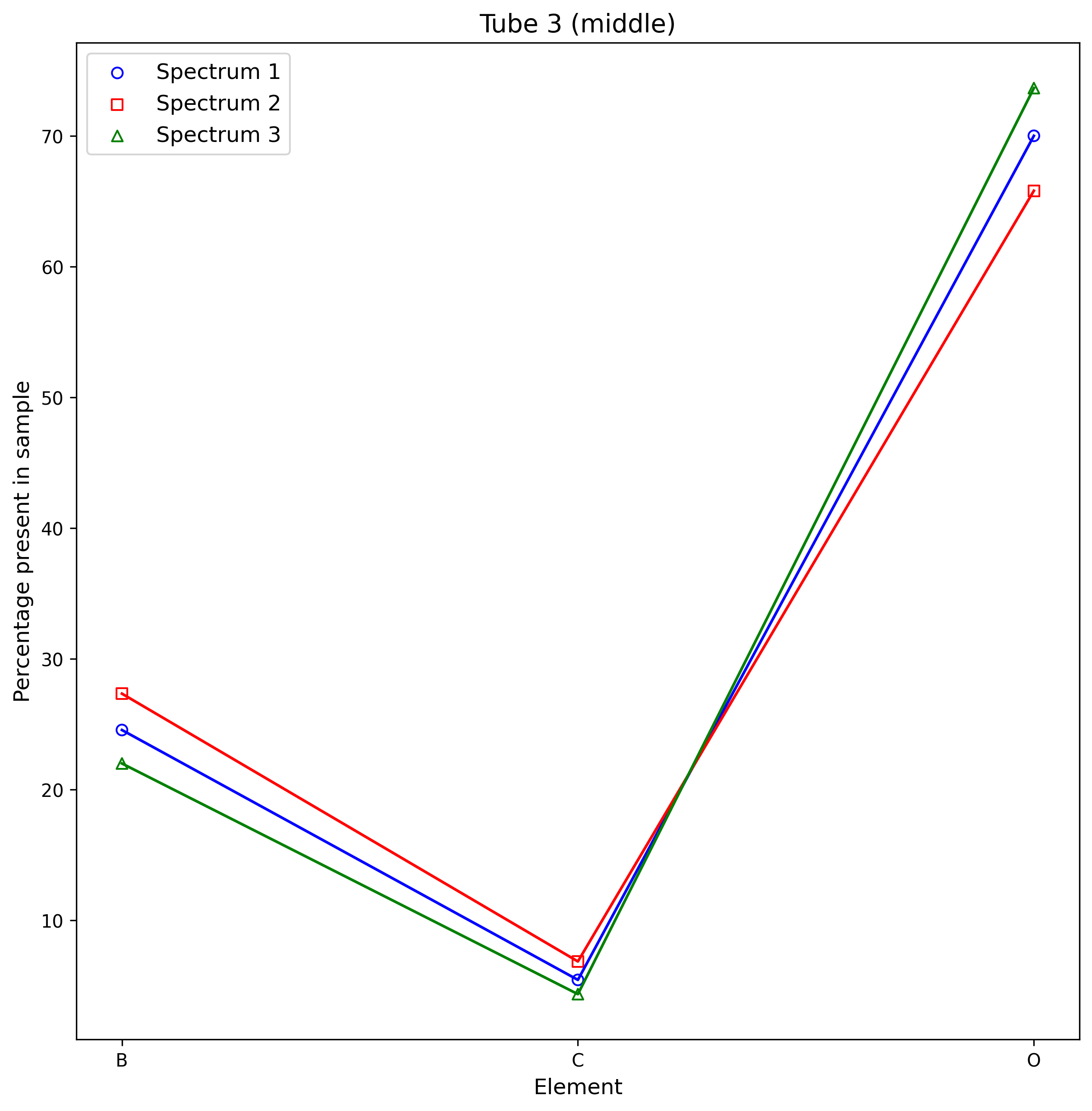}
\noindent\includegraphics[width=0.4\textwidth]{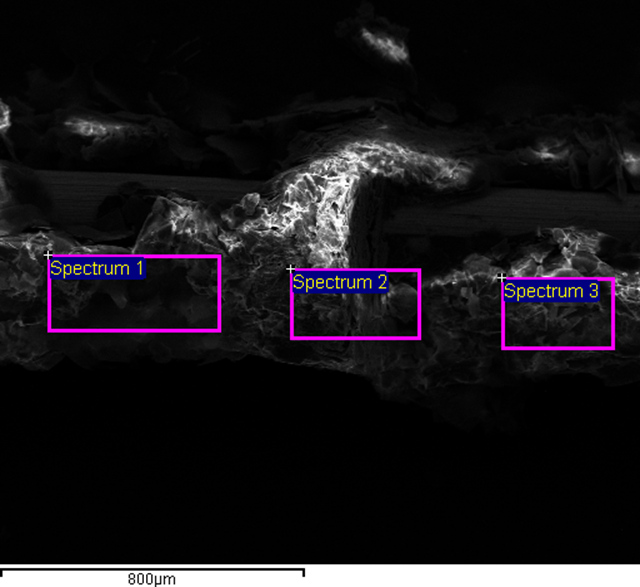}\\
\noindent\includegraphics[width=0.38\textwidth]{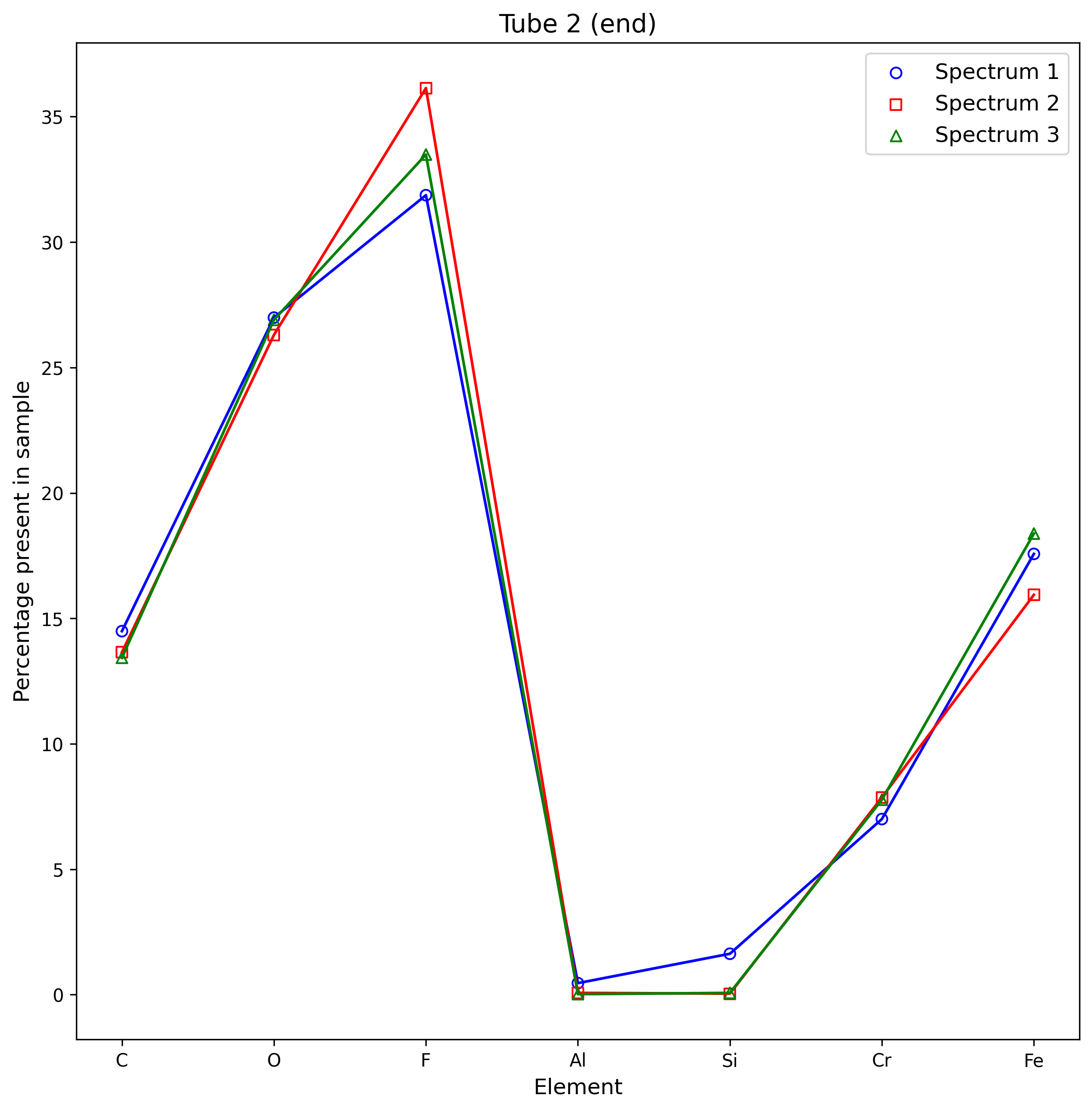}
\noindent\includegraphics[width=0.4\textwidth]{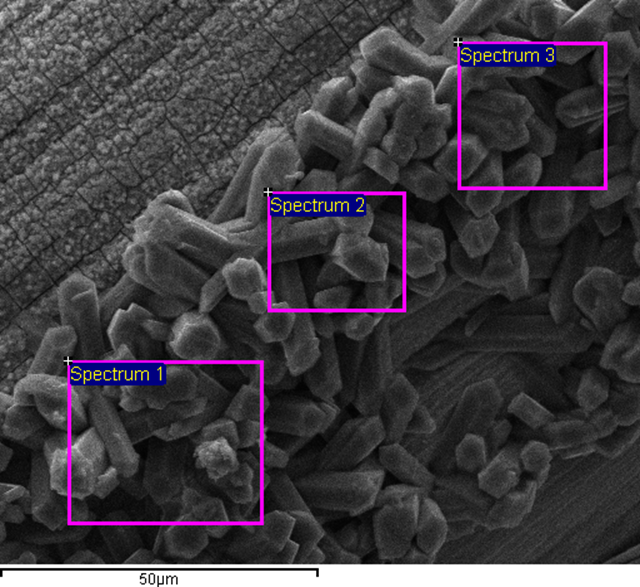}
\caption{Results of the composition analysis of the crystal structures found on several {central wires}. }
\label{Fig:tube_analysis}
\end{center}
\end{figure}

To identify the crystal structures, the samples were analyzed using energy-dispersive X-ray spectroscopy (EDS). However, this method cannot readily identify light elements in a sample due to the lack of characteristic X-ray emission. Results of the EDS analysis is presented in Fig. \ref{Fig:tube_analysis} for three different tubes (other samples were very similar), with the composition results shown in the left panels, and the magnified samples in the right panels.

For Tube \#1 (top panels), spectrum \#5 shows the results of a relatively ``clean" {central wire} consisting of $>80\%$ Tungsten (the green curve). Analysis of the larger crystal structures covering parts of both Tube \#1 and \#3 (middle panel) showed high concentrations of B and O. BF$_3$ gas is extremely reactive with water and/or moisture, leading to the following reaction

\begin{equation}
4 \mathrm{BF}_3 + 2 \mathrm{H}_2\mathrm{O} \rightarrow 3 \mathrm{HBF}_4 + \mathrm{B(OH)}_3
\end{equation}

which forms two acid compounds. Visually, the BF$_3$ gas turns white on contact with moisture. The analysis for Tubes \#1 and \#3 seem to indicate the possible presence of boric acid (B(OH)$_3$) crystals on the {central wire}. We can therefore conclude that these tubes malfunctioned due to a broken seal allowing BF$_3$ gas to escape and moisture to penetrate the tube.\\

The smaller crystal structure observed on the {central wire} of Tube \#2 (bottom panels), however, show large concentration of F, which is a by-product of neutron capture process

\begin{equation}
^{10} \mathrm{BF}_3 + \mathrm{n} \rightarrow \, ^7 \mathrm{Li}^{+} + \alpha^{+2} + 3\mathrm{F}^{-}
\end{equation}

The finer crystal structures present in the samples {could therefore be} LiF crystals formed during the neutron capture process, keeping in mind that the EDS technique cannot measure the lighter Li element. {However, this can also be F compounds as F is very reactive and could bind with and form compounds with a range of elements present in the tubes.} \\

{The results presented in this section seem to confirm the statement by \citet{Bieberetal2007} that material formed during neutron detection process can ``plate out" onto the central wire and decrease the detection efficiency of the tube. This is most likely BF$_3$ compounds released in the cascade, with trace amounts of material formed directly in the neutron capture reaction.}

%-------------------------------------------------------------

%-------------------------------------------------------------
\newpage 

\bibliographystyle{model5-names}
\biboptions{authoryear}
\bibliography{mybibfile}

\begin{thebibliography}{30}
\expandafter\ifx\csname natexlab\endcsname\relax\def\natexlab#1{#1}\fi
\providecommand{\url}[1]{\texttt{#1}}
\providecommand{\href}[2]{#2}
\providecommand{\path}[1]{#1}
\providecommand{\DOIprefix}{doi:}
\providecommand{\ArXivprefix}{arXiv:}
\providecommand{\URLprefix}{URL: }
\providecommand{\Pubmedprefix}{pmid:}
\providecommand{\doi}[1]{\href{http://dx.doi.org/#1}{\path{#1}}}
\providecommand{\Pubmed}[1]{\href{pmid:#1}{\path{#1}}}
\providecommand{\bibinfo}[2]{#2}
\ifx\xfnm\relax \def\xfnm[#1]{\unskip,\space#1}\fi
%Type = Article
\bibitem[{{Balabin} et~al.(2011){Balabin}, {Gvozdevsk}, {Maurchev}, {Vashenyuk}
  \& {Dzhappuev}}]{balabin2011}
\bibinfo{author}{{Balabin}, Y.~V.}, \bibinfo{author}{{Gvozdevsk}, B.~B.},
  \bibinfo{author}{{Maurchev}, E.~A.}, \bibinfo{author}{{Vashenyuk}, E.~V.}, \&
  \bibinfo{author}{{Dzhappuev}, D.~D.} (\bibinfo{year}{2011}).
\newblock \bibinfo{title}{{Fine structure of neutron multiplicity on neutron
  monitors}}.
\newblock {\it \bibinfo{journal}{Astrophysics and Space Sciences
  Transactions}\/},  {\it \bibinfo{volume}{7}\/}\bibinfo{issue}{(3)},
  \bibinfo{pages}{283--286}. \DOIprefix\doi{10.5194/astra-7-283-2011}.
%Type = Article
\bibitem[{{Bieber} et~al.(2007){Bieber}, {Clem}, {Desilets}, {Evenson}, {Lal},
  {Lopate} \& {Pyle}}]{Bieberetal2007}
\bibinfo{author}{{Bieber}, J.~W.}, \bibinfo{author}{{Clem}, J.},
  \bibinfo{author}{{Desilets}, D.}, \bibinfo{author}{{Evenson}, P.},
  \bibinfo{author}{{Lal}, D.}, \bibinfo{author}{{Lopate}, C.}, \&
  \bibinfo{author}{{Pyle}, R.} (\bibinfo{year}{2007}).
\newblock \bibinfo{title}{{Long-term decline of South Pole neutron rates}}.
\newblock {\it \bibinfo{journal}{Journal of Geophysical Research (Space
  Physics)}\/},  {\it \bibinfo{volume}{112}\/}\bibinfo{issue}{(A12)},
  \bibinfo{pages}{A12102}. \DOIprefix\doi{10.1029/2006JA011894}.
%Type = Article
\bibitem[{{Bieber} et~al.(2004){Bieber}, {Clem}, {Duldig}, {Evenson}, {Humble}
  \& {Pyle}}]{bieberetal2004}
\bibinfo{author}{{Bieber}, J.~W.}, \bibinfo{author}{{Clem}, J.~M.},
  \bibinfo{author}{{Duldig}, M.~L.}, \bibinfo{author}{{Evenson}, P.~A.},
  \bibinfo{author}{{Humble}, J.~E.}, \& \bibinfo{author}{{Pyle}, R.}
  (\bibinfo{year}{2004}).
\newblock \bibinfo{title}{{Latitude survey observations of neutron monitor
  multiplicity}}.
\newblock {\it \bibinfo{journal}{Journal of Geophysical Research (Space
  Physics)}\/},  {\it \bibinfo{volume}{109}\/}\bibinfo{issue}{(A12)},
  \bibinfo{pages}{A12106}. \DOIprefix\doi{10.1029/2004JA010493}.
%Type = Inbook
\bibitem[{B{\"u}tikofer(2018)}]{Butikofer2018}
\bibinfo{author}{B{\"u}tikofer, R.} (\bibinfo{year}{2018}).
\newblock \bibinfo{title}{Ground-based measurements of energetic particles by
  neutron monitors}.
\newblock In \bibinfo{editor}{O.~E. {Malandraki}}, \& \bibinfo{editor}{N.~B.
  {Crosby}} (Eds.), {\it \bibinfo{booktitle}{Solar Particle Radiation Storms
  Forecasting and Analysis: The HESPERIA HORIZON 2020 Project and Beyond}\/}
  (pp. \bibinfo{pages}{95--111}).
\newblock \bibinfo{publisher}{Springer International Publishing}.
%Type = Article
\bibitem[{{Caballero-Lopez}(2016)}]{rogelio2016}
\bibinfo{author}{{Caballero-Lopez}, R.~A.} (\bibinfo{year}{2016}).
\newblock \bibinfo{title}{{An estimation of the yield and response functions
  for the mini neutron monitor}}.
\newblock {\it \bibinfo{journal}{Journal of Geophysical Research (Space
  Physics)}\/},  {\it \bibinfo{volume}{121}\/}\bibinfo{issue}{(8)},
  \bibinfo{pages}{7461--7469}. \DOIprefix\doi{10.1002/2016JA022690}.
%Type = Article
\bibitem[{{Carmichael} et~al.(1968){Carmichael}, {Bercovitch}, {Shea},
  {Magidin} \& {Peterson}}]{carmichael1968}
\bibinfo{author}{{Carmichael}, H.}, \bibinfo{author}{{Bercovitch}, M.},
  \bibinfo{author}{{Shea}, M.~A.}, \bibinfo{author}{{Magidin}, M.}, \&
  \bibinfo{author}{{Peterson}, R.~W.} (\bibinfo{year}{1968}).
\newblock \bibinfo{title}{{Attenuation of neutron monitor radiation in the
  atmosphere}}.
\newblock {\it \bibinfo{journal}{Canadian Journal of Physics Supplement}\/},
  {\it \bibinfo{volume}{46}\/}, \bibinfo{pages}{S1006--S1013}.
  \DOIprefix\doi{10.1139/p68-405}.
%Type = Article
\bibitem[{{Clem} \& {Dorman}(2000)}]{clemdorman2000}
\bibinfo{author}{{Clem}, J.~M.}, \& \bibinfo{author}{{Dorman}, L.~I.}
  (\bibinfo{year}{2000}).
\newblock \bibinfo{title}{{Neutron Monitor Response Functions}}.
\newblock {\it \bibinfo{journal}{Space Science Reviews}\/},  {\it
  \bibinfo{volume}{93}\/}, \bibinfo{pages}{335--359}.
  \DOIprefix\doi{10.1023/A:1026508915269}.
%Type = Article
\bibitem[{{Desorgher} et~al.(2009){Desorgher}, {Kudela}, {Fl{\"u}ckiger},
  {B{\"u}tikofer}, {Storini} \& {Kalegaev}}]{Desorgher-etal-2009}
\bibinfo{author}{{Desorgher}, L.}, \bibinfo{author}{{Kudela}, K.},
  \bibinfo{author}{{Fl{\"u}ckiger}, E.}, \bibinfo{author}{{B{\"u}tikofer}, R.},
  \bibinfo{author}{{Storini}, M.}, \& \bibinfo{author}{{Kalegaev}, V.}
  (\bibinfo{year}{2009}).
\newblock \bibinfo{title}{{Comparison of Earth's magnetospheric magnetic field
  models in the context of cosmic ray physics}}.
\newblock {\it \bibinfo{journal}{Acta Geophysica}\/},  {\it
  \bibinfo{volume}{57}\/}\bibinfo{issue}{(1)}, \bibinfo{pages}{75--87}.
  \DOIprefix\doi{10.2478/s11600-008-0065-3}.
%Type = Book
\bibitem[{{Dorman}(2004)}]{dorman2004}
\bibinfo{author}{{Dorman}, L.~I.} (\bibinfo{year}{2004}).
\newblock {\it \bibinfo{title}{{Cosmic Rays in the Earth's Atmosphere and
  Underground}}\/}.
\newblock \bibinfo{address}{Netherlands}: \bibinfo{publisher}{Springer}.
%Type = Inproceedings
\bibitem[{{Evenson} et~al.(2005){Evenson}, {Bieber}, {Clem} \&
  {Pyle}}]{evensonetal2005}
\bibinfo{author}{{Evenson}, P.}, \bibinfo{author}{{Bieber}, J.~W.},
  \bibinfo{author}{{Clem}, J.}, \& \bibinfo{author}{{Pyle}, R.}
  (\bibinfo{year}{2005}).
\newblock \bibinfo{title}{{Neutron Monitor Temperature Coefficients:
  Measurements for BF3 and 3He Counter Tubes}}.
\newblock In {\it \bibinfo{booktitle}{29th International Cosmic Ray Conference
  (ICRC29), Volume 2}\/} (p. \bibinfo{pages}{485}).
\newblock volume~\bibinfo{volume}{2} of {\it \bibinfo{series}{International
  Cosmic Ray Conference}\/}.
%Type = Article
\bibitem[{{Fowler}(1963)}]{fowler1963}
\bibinfo{author}{{Fowler}, I.~L.} (\bibinfo{year}{1963}).
\newblock \bibinfo{title}{{Very Large Boron Trifluoride Proportional
  Counters}}.
\newblock {\it \bibinfo{journal}{Review of Scientific Instruments}\/},  {\it
  \bibinfo{volume}{34}\/}\bibinfo{issue}{(7)}, \bibinfo{pages}{731--739}.
  \DOIprefix\doi{10.1063/1.1718559}.
%Type = Article
\bibitem[{Harris et~al.(2020)Harris, Millman, van~der Walt, Gommers, Virtanen,
  Cournapeau, Wieser, Taylor, Berg, Smith, Kern, Picus, Hoyer, van Kerkwijk,
  Brett, Haldane, del Río, Wiebe, Peterson, Gérard-Marchant, Sheppard, Reddy,
  Weckesser, Abbasi, Gohlke \& Oliphant}]{harrisetal2020}
\bibinfo{author}{Harris, C.~R.}, \bibinfo{author}{Millman, K.~J.},
  \bibinfo{author}{van~der Walt, S.~J.}, \bibinfo{author}{Gommers, R.},
  \bibinfo{author}{Virtanen, P.}, \bibinfo{author}{Cournapeau, D.},
  \bibinfo{author}{Wieser, E.}, \bibinfo{author}{Taylor, J.},
  \bibinfo{author}{Berg, S.}, \bibinfo{author}{Smith, N.~J.},
  \bibinfo{author}{Kern, R.}, \bibinfo{author}{Picus, M.},
  \bibinfo{author}{Hoyer, S.}, \bibinfo{author}{van Kerkwijk, M.~H.},
  \bibinfo{author}{Brett, M.}, \bibinfo{author}{Haldane, A.},
  \bibinfo{author}{del Río, J.~F.}, \bibinfo{author}{Wiebe, M.},
  \bibinfo{author}{Peterson, P.}, \bibinfo{author}{Gérard-Marchant, P.},
  \bibinfo{author}{Sheppard, K.}, \bibinfo{author}{Reddy, T.},
  \bibinfo{author}{Weckesser, W.}, \bibinfo{author}{Abbasi, H.},
  \bibinfo{author}{Gohlke, C.}, \& \bibinfo{author}{Oliphant, T.~E.}
  (\bibinfo{year}{2020}).
\newblock \bibinfo{title}{{Array programming with NumPy}}.
\newblock {\it \bibinfo{journal}{Nature}\/},  {\it \bibinfo{volume}{585}\/},
  \bibinfo{pages}{357--362}. \DOIprefix\doi{10.1038/s41586-020-2649-2}.
%Type = Article
\bibitem[{{Hatton} \& {Carimichael}(1964)}]{HattonCarmicheal1964}
\bibinfo{author}{{Hatton}, C.~J.}, \& \bibinfo{author}{{Carimichael}, H.}
  (\bibinfo{year}{1964}).
\newblock \bibinfo{title}{{Experimental Investigation of the NM-64 Neutron
  Monitor}}.
\newblock {\it \bibinfo{journal}{Canadian Journal of Physics}\/},  {\it
  \bibinfo{volume}{42}\/}\bibinfo{issue}{(12)}, \bibinfo{pages}{2443--2472}.
  \DOIprefix\doi{10.1139/p64-222}.
%Type = Article
\bibitem[{Hunter(2007)}]{Matplotlib-2007}
\bibinfo{author}{Hunter, J.~D.} (\bibinfo{year}{2007}).
\newblock \bibinfo{title}{Matplotlib: A 2d graphics environment}.
\newblock {\it \bibinfo{journal}{Computing in Science \& Engineering}\/},  {\it
  \bibinfo{volume}{9}\/}\bibinfo{issue}{(3)}, \bibinfo{pages}{90--95}.
  \DOIprefix\doi{10.1109/MCSE.2007.55}.
%Type = Article
\bibitem[{{Kato} et~al.(2021){Kato}, {Kihara}, {Ko}, {Kadokura}, {Kataoka},
  {Evenson}, {Uchida}, {Kaimi}, {Nakamura}, {Uchida}, {Murase} \&
  {Munakata}}]{Katoetal2001}
\bibinfo{author}{{Kato}, C.}, \bibinfo{author}{{Kihara}, W.},
  \bibinfo{author}{{Ko}, Y.}, \bibinfo{author}{{Kadokura}, A.},
  \bibinfo{author}{{Kataoka}, R.}, \bibinfo{author}{{Evenson}, P.},
  \bibinfo{author}{{Uchida}, S.}, \bibinfo{author}{{Kaimi}, S.},
  \bibinfo{author}{{Nakamura}, Y.}, \bibinfo{author}{{Uchida}, H.~A.},
  \bibinfo{author}{{Murase}, K.}, \& \bibinfo{author}{{Munakata}, K.}
  (\bibinfo{year}{2021}).
\newblock \bibinfo{title}{{New cosmic ray observations at Syowa Station in the
  Antarctic for space weather study}}.
\newblock {\it \bibinfo{journal}{arXiv e-prints}\/},  (p.
  \bibinfo{pages}{arXiv:2101.09887}).
  \href{http://arxiv.org/abs/2101.09887}{\tt arXiv:2101.09887}.
%Type = Book
\bibitem[{Knoll(2010)}]{knoll12010}
\bibinfo{author}{Knoll, G.~F.} (\bibinfo{year}{2010}).
\newblock {\it \bibinfo{title}{{Radiation Detection and Measurement, 4th
  ed.}}\/}.
\newblock \bibinfo{address}{New York, NY}: \bibinfo{publisher}{Wiley}.
%Type = Article
\bibitem[{{Kr{\"u}ger} et~al.(2008){Kr{\"u}ger}, {Moraal}, {Bieber}, {Clem},
  {Evenson}, {Pyle}, {Duldig} \& {Humble}}]{krugeretal2008}
\bibinfo{author}{{Kr{\"u}ger}, H.}, \bibinfo{author}{{Moraal}, H.},
  \bibinfo{author}{{Bieber}, J.~W.}, \bibinfo{author}{{Clem}, J.~M.},
  \bibinfo{author}{{Evenson}, P.~A.}, \bibinfo{author}{{Pyle}, K.~R.},
  \bibinfo{author}{{Duldig}, M.~L.}, \& \bibinfo{author}{{Humble}, J.~E.}
  (\bibinfo{year}{2008}).
\newblock \bibinfo{title}{{A calibration neutron monitor: Energy response and
  instrumental temperature sensitivity}}.
\newblock {\it \bibinfo{journal}{Journal of Geophysical Research (Space
  Physics)}\/},  {\it \bibinfo{volume}{113}\/}\bibinfo{issue}{(A8)},
  \bibinfo{pages}{A08101}. \DOIprefix\doi{10.1029/2008JA013229}.
%Type = Inproceedings
\bibitem[{Kr{\"u}ger et~al.(2017)Kr{\"u}ger, Kr{\"u}ger, Kr{\"u}ger, Diedericks
  \& Malan}]{Krugeretal2017}
\bibinfo{author}{Kr{\"u}ger, P.}, \bibinfo{author}{Kr{\"u}ger, H.},
  \bibinfo{author}{Kr{\"u}ger, H.}, \bibinfo{author}{Diedericks, C.}, \&
  \bibinfo{author}{Malan, D.} (\bibinfo{year}{2017}).
\newblock \bibinfo{title}{{New, affordable, open-hardware neutron monitor
  electronics}}.
\newblock In {\it \bibinfo{booktitle}{Proceedings of 35th International Cosmic
  Ray Conference {\textemdash} PoS(ICRC2017)}\/} (p. \bibinfo{pages}{054}).
\newblock volume \bibinfo{volume}{301}.
\newblock \DOIprefix\doi{10.22323/1.301.0054}.
%Type = Article
\bibitem[{Malan \& Moraal(2002)}]{malanmoraal2002}
\bibinfo{author}{Malan, N.}, \& \bibinfo{author}{Moraal, H.}
  (\bibinfo{year}{2002}).
\newblock \bibinfo{title}{{The effect of wind on pressure correction of the
  SANAE neutron monitor counting rate}}.
\newblock {\it \bibinfo{journal}{South African Journal of Science}\/},  {\it
  \bibinfo{volume}{98(5)}\/}, \bibinfo{pages}{278--281}.
%Type = Article
\bibitem[{{Mangeard} et~al.(2016){Mangeard}, {Ruffolo}, {S{\'a}iz},
  {Nuntiyakul}, {Bieber}, {Clem}, {Evenson}, {Pyle}, {Duldig} \&
  {Humble}}]{mangeard2016}
\bibinfo{author}{{Mangeard}, P.~S.}, \bibinfo{author}{{Ruffolo}, D.},
  \bibinfo{author}{{S{\'a}iz}, A.}, \bibinfo{author}{{Nuntiyakul}, W.},
  \bibinfo{author}{{Bieber}, J.~W.}, \bibinfo{author}{{Clem}, J.},
  \bibinfo{author}{{Evenson}, P.}, \bibinfo{author}{{Pyle}, R.},
  \bibinfo{author}{{Duldig}, M.~L.}, \& \bibinfo{author}{{Humble}, J.~E.}
  (\bibinfo{year}{2016}).
\newblock \bibinfo{title}{{Dependence of the neutron monitor count rate and
  time delay distribution on the rigidity spectrum of primary cosmic rays}}.
\newblock {\it \bibinfo{journal}{Journal of Geophysical Research (Space
  Physics)}\/},  {\it \bibinfo{volume}{121}\/}\bibinfo{issue}{(12)},
  \bibinfo{pages}{11,620--11,636}. \DOIprefix\doi{10.1002/2016JA023515}.
%Type = Article
\bibitem[{{Medina} et~al.(2013){Medina}, {Blanco}, {Garc{\'\i}a},
  {G{\'o}mez-Herrero}, {Catal{\'a}n}, {Garc{\'\i}a}, {Hidalgo}, {Meziat},
  {Prieto}, {Rodr{\'\i}guez-Pacheco} \& {S{\'a}nchez}}]{medina2013}
\bibinfo{author}{{Medina}, J.}, \bibinfo{author}{{Blanco}, J.~J.},
  \bibinfo{author}{{Garc{\'\i}a}, O.}, \bibinfo{author}{{G{\'o}mez-Herrero},
  R.}, \bibinfo{author}{{Catal{\'a}n}, E.~J.}, \bibinfo{author}{{Garc{\'\i}a},
  I.}, \bibinfo{author}{{Hidalgo}, M.~A.}, \bibinfo{author}{{Meziat}, D.},
  \bibinfo{author}{{Prieto}, M.}, \bibinfo{author}{{Rodr{\'\i}guez-Pacheco},
  J.}, \& \bibinfo{author}{{S{\'a}nchez}, S.} (\bibinfo{year}{2013}).
\newblock \bibinfo{title}{{Castilla-La Mancha neutron monitor}}.
\newblock {\it \bibinfo{journal}{Nuclear Instruments and Methods in Physics
  Research A}\/},  {\it \bibinfo{volume}{727}\/}, \bibinfo{pages}{97--103}.
  \DOIprefix\doi{10.1016/j.nima.2013.06.028}.
%Type = Article
\bibitem[{{Mishev} \& {Usoskin}(2020)}]{Mishev_Usoskin_2020}
\bibinfo{author}{{Mishev}, A.}, \& \bibinfo{author}{{Usoskin}, I.}
  (\bibinfo{year}{2020}).
\newblock \bibinfo{title}{{Current status and possible extension of the global
  neutron monitor network}}.
\newblock {\it \bibinfo{journal}{Journal of Space Weather and Space
  Climate}\/},  {\it \bibinfo{volume}{10}\/}, \bibinfo{pages}{17}.
  \DOIprefix\doi{10.1051/swsc/2020020}.
  \href{http://arxiv.org/abs/2005.12621}{\tt arXiv:2005.12621}.
%Type = Article
\bibitem[{{Mishev} et~al.(2020){Mishev}, {Koldobskiy}, {Kovaltsov}, {Gil} \&
  {Usoskin}}]{mishevetal2020}
\bibinfo{author}{{Mishev}, A.~L.}, \bibinfo{author}{{Koldobskiy}, S.~A.},
  \bibinfo{author}{{Kovaltsov}, G.~A.}, \bibinfo{author}{{Gil}, A.}, \&
  \bibinfo{author}{{Usoskin}, I.~G.} (\bibinfo{year}{2020}).
\newblock \bibinfo{title}{{Updated Neutron-Monitor Yield Function: Bridging
  Between In Situ and Ground-Based Cosmic Ray Measurements}}.
\newblock {\it \bibinfo{journal}{Journal of Geophysical Research (Space
  Physics)}\/},  {\it \bibinfo{volume}{125}\/}\bibinfo{issue}{(2)},
  \bibinfo{pages}{e27433}. \DOIprefix\doi{10.1029/2019JA027433}.
%Type = Inproceedings
\bibitem[{{Moraal} et~al.(2011){Moraal}, {Stoker}, {Humble} \&
  {Mans}}]{moraaletal2011}
\bibinfo{author}{{Moraal}, H.}, \bibinfo{author}{{Stoker}, P.},
  \bibinfo{author}{{Humble}, J.}, \& \bibinfo{author}{{Mans}, A.}
  (\bibinfo{year}{2011}).
\newblock \bibinfo{title}{{Long-term Data Records of South African Neutron
  Monitors}}.
\newblock In {\it \bibinfo{booktitle}{International Cosmic Ray Conference}\/}
  (p. \bibinfo{pages}{187}).
\newblock volume~\bibinfo{volume}{11} of {\it \bibinfo{series}{International
  Cosmic Ray Conference}\/}.
\newblock \DOIprefix\doi{10.7529/ICRC2011/V11/0475}.
%Type = Article
\bibitem[{{Paschalis} et~al.(2013){Paschalis}, {Mavromichalaki}, {Yanke},
  {Belov}, {Eroshenko}, {Gerontidou} \& {Koutroumpi}}]{Paschalisetal2013}
\bibinfo{author}{{Paschalis}, P.}, \bibinfo{author}{{Mavromichalaki}, H.},
  \bibinfo{author}{{Yanke}, V.}, \bibinfo{author}{{Belov}, A.},
  \bibinfo{author}{{Eroshenko}, E.}, \bibinfo{author}{{Gerontidou}, M.}, \&
  \bibinfo{author}{{Koutroumpi}, I.} (\bibinfo{year}{2013}).
\newblock \bibinfo{title}{{Online application for the barometric coefficient
  calculation of the NMDB stations}}.
\newblock {\it \bibinfo{journal}{New Astronomy}\/},  {\it
  \bibinfo{volume}{19}\/}, \bibinfo{pages}{10--18}.
  \DOIprefix\doi{10.1016/j.newast.2012.08.003}.
%Type = Article
\bibitem[{{Ruffolo} et~al.(2016){Ruffolo}, {S{\'a}iz}, {Mangeard}, {Kamyan},
  {Muangha}, {Nutaro}, {Sumran}, {Chaiwattana}, {Gasiprong}, {Channok},
  {Wuttiya}, {Rujiwarodom}, {Tooprakai}, {Asavapibhop}, {Bieber}, {Clem},
  {Evenson} \& {Munakata}}]{ruffoloetal2016}
\bibinfo{author}{{Ruffolo}, D.}, \bibinfo{author}{{S{\'a}iz}, A.},
  \bibinfo{author}{{Mangeard}, P.~S.}, \bibinfo{author}{{Kamyan}, N.},
  \bibinfo{author}{{Muangha}, P.}, \bibinfo{author}{{Nutaro}, T.},
  \bibinfo{author}{{Sumran}, S.}, \bibinfo{author}{{Chaiwattana}, C.},
  \bibinfo{author}{{Gasiprong}, N.}, \bibinfo{author}{{Channok}, C.},
  \bibinfo{author}{{Wuttiya}, C.}, \bibinfo{author}{{Rujiwarodom}, M.},
  \bibinfo{author}{{Tooprakai}, P.}, \bibinfo{author}{{Asavapibhop}, B.},
  \bibinfo{author}{{Bieber}, J.~W.}, \bibinfo{author}{{Clem}, J.},
  \bibinfo{author}{{Evenson}, P.}, \& \bibinfo{author}{{Munakata}, K.}
  (\bibinfo{year}{2016}).
\newblock \bibinfo{title}{{Monitoring Short-term Cosmic-ray Spectral Variations
  Using Neutron Monitor Time-delay Measurements}}.
\newblock {\it \bibinfo{journal}{Astrophysical Journal}\/},  {\it
  \bibinfo{volume}{817}\/}\bibinfo{issue}{(1)}, \bibinfo{pages}{38}.
  \DOIprefix\doi{10.3847/0004-637X/817/1/38}.
%Type = Inproceedings
\bibitem[{S\'aiz et~al.(2019)S\'aiz, Mitthumsiri, Ruffolo, Evenson \&
  Nutaro}]{saiz2019}
\bibinfo{author}{S\'aiz, A.}, \bibinfo{author}{Mitthumsiri, W.},
  \bibinfo{author}{Ruffolo, D.}, \bibinfo{author}{Evenson, P.}, \&
  \bibinfo{author}{Nutaro, T.} (\bibinfo{year}{2019}).
\newblock \bibinfo{title}{{Detecting Single and Multiple Atmospheric
  Secondaries in an 18NM64}}.
\newblock In {\it \bibinfo{booktitle}{Proceedings of 36th International Cosmic
  Ray Conference {\textemdash} PoS(ICRC2019)}\/} (p. \bibinfo{pages}{1145}).
\newblock volume \bibinfo{volume}{358}.
\newblock \DOIprefix\doi{10.22323/1.358.1145}.
%Type = Article
\bibitem[{{Simil{\"a}} et~al.(2021){Simil{\"a}}, {Usoskin}, {Poluianov},
  {Mishev}, {Kovaltsov} \& {Toit Strauss}}]{Similia2021}
\bibinfo{author}{{Simil{\"a}}, M.}, \bibinfo{author}{{Usoskin}, I.},
  \bibinfo{author}{{Poluianov}, S.}, \bibinfo{author}{{Mishev}, A.},
  \bibinfo{author}{{Kovaltsov}, G.~A.}, \& \bibinfo{author}{{Toit Strauss}, D.}
  (\bibinfo{year}{2021}).
\newblock \bibinfo{title}{{High-altitude polar NM with the new DAQ system as a
  tool to study details of the cosmic-ray induced nucleonic cascade}}.
\newblock {\it \bibinfo{journal}{arXiv e-prints}\/},  (p.
  \bibinfo{pages}{arXiv:2104.04727}).
  \href{http://arxiv.org/abs/2104.04727}{\tt arXiv:2104.04727}.
%Type = Article
\bibitem[{{Simpson}(2000)}]{simpson2000}
\bibinfo{author}{{Simpson}, J.~A.} (\bibinfo{year}{2000}).
\newblock \bibinfo{title}{{The Cosmic Ray Nucleonic Component: The Invention
  and Scientific Uses of the Neutron Monitor - (Keynote Lecture)}}.
\newblock {\it \bibinfo{journal}{Space Science Reviews}\/},  {\it
  \bibinfo{volume}{93}\/}, \bibinfo{pages}{11--32}.
  \DOIprefix\doi{10.1023/A:1026567706183}.
%Type = Article
\bibitem[{{Strauss} et~al.(2020){Strauss}, {Poluianov}, {van der Merwe},
  {Kr{\"u}ger}, {Diedericks}, {Kr{\"u}ger}, {Usoskin}, {Heber}, {Nndanganeni},
  {Blanco-{\'A}valos}, {Garc{\'\i}a-Tejedor}, {Herbst}, {Caballero-Lopez},
  {Moloto}, {Lara}, {Walter}, {Giday} \& {Traversi}}]{Straussetal2020}
\bibinfo{author}{{Strauss}, D.~T.}, \bibinfo{author}{{Poluianov}, S.},
  \bibinfo{author}{{van der Merwe}, C.}, \bibinfo{author}{{Kr{\"u}ger}, H.},
  \bibinfo{author}{{Diedericks}, C.}, \bibinfo{author}{{Kr{\"u}ger}, H.},
  \bibinfo{author}{{Usoskin}, I.}, \bibinfo{author}{{Heber}, B.},
  \bibinfo{author}{{Nndanganeni}, R.}, \bibinfo{author}{{Blanco-{\'A}valos},
  J.}, \bibinfo{author}{{Garc{\'\i}a-Tejedor}, I.}, \bibinfo{author}{{Herbst},
  K.}, \bibinfo{author}{{Caballero-Lopez}, R.}, \bibinfo{author}{{Moloto}, K.},
  \bibinfo{author}{{Lara}, A.}, \bibinfo{author}{{Walter}, M.},
  \bibinfo{author}{{Giday}, N.~M.}, \& \bibinfo{author}{{Traversi}, R.}
  (\bibinfo{year}{2020}).
\newblock \bibinfo{title}{{The mini-neutron monitor: a new approach in neutron
  monitor design}}.
\newblock {\it \bibinfo{journal}{Journal of Space Weather and Space
  Climate}\/},  {\it \bibinfo{volume}{10}\/}, \bibinfo{pages}{39}.
  \DOIprefix\doi{10.1051/swsc/2020038}.

\end{thebibliography}

\end{document}